\PassOptionsToPackage{prologue,table,xcdraw}{xcolor}
\documentclass[sigconf,authorversion]{acmart}
\usepackage{acronym}
\usepackage{acmart-taps}
\usepackage{amsmath}
\usepackage{amsfonts}
\usepackage[capitalize,noabbrev,english]{cleveref}
\usepackage{framed}
\usepackage{graphicx}
\usepackage{multirow}
\usepackage{siunitx}
\usepackage{subcaption}
\usepackage{textcomp}
\usepackage{xcolor}
\usepackage{xspace}

\definecolor{mypurple}{HTML}{C4C5E6}
\definecolor{myblue}{HTML}{5E62B8}
\definecolor{mybeige}{HTML}{DCC0B0}
\definecolor{mylightbeige}{HTML}{F6DDCA}
\definecolor{shadecolor}{RGB}{247, 243, 239}
\definecolor{aa}{RGB}{239, 241, 247}
\definecolor{ab}{RGB}{222, 224, 228}
\definecolor{ba}{RGB}{247, 243, 239}
\definecolor{bb}{RGB}{242, 233, 224}
\definecolor{cheatmap}{RGB}{220, 192, 176} 

\newlength{\comicwidth}
\newlength{\comicmargin}

\newcolumntype{R}{>{\raggedleft\arraybackslash}X}
\newcolumntype{L}[1]{>{\raggedright\arraybackslash}p{#1}}
\newcolumntype{C}[1]{>{\centering\arraybackslash}p{#1}}

\DeclareSIUnit{\year}{yr}
\DeclareSIUnit{\pp}{p.p.}

\def\EUR{\text{\texteuro}}

\newcommand{\appcategory}[1]{\colorbox{ba}{\textit{#1}}}
\newcommand{\associationstats}[4]{$(\chi^2(#1) = \num{#2}, p < \num{#3}, V = \num{#4})$}
\newcommand{\captionheadline}[1]{\textbf{\textit{#1.}}}
\newcommand{\cc}[1]{\cellcolor{#1}}

\newcommand{\designcriterion}[1]{\colorbox{aa}{\textit{#1}}\xspace}
\newcommand{\diffrank}[1]{(\ensuremath{#1} ranks)\xspace}
\newcommand{\gapfish}{GapFish\xspace}
\newcommand{\hmc}[1]{\cellcolor{cheatmap!#1}}
\newcommand{\hmct}[1]{\cellcolor{cheatmap!#1} #1~\%}
\newcommand{\hmctb}[1]{\cellcolor{cheatmap!#1} \textbf{#1~\%}}
\newcommand{\hourmin}[2]{\qty{#1}{\hour}\xspace\qty{#2}{\minute}}

\newcommand{\minsec}[2]{\qty{#1}{\minute}\xspace\qty{#2}{\second}}
\newcommand{\medianrank}[1]{(median Rank #1)\xspace}
\newcommand{\mytablemark}[1]{$^#1$}
\newcommand{\mytablenote}[4]{\multicolumn{#2}{p{#3}}{\scriptsize{$^#1$ #4}}\\}
\newcommand{\npartials}{\num{2187}\xspace}
\newcommand{\ncomplete}{\num{986}\xspace}
\newcommand{\nfinal}{\num{857}\xspace}

\newcommand{\rccc}[2]{\rowcolor{#1}\cellcolor{#2}}
\newcommand{\researchquestion}[1]{\textbf{#1}\xspace}

\newcommand{\smalln}[1]{($n=\num{#1}$)\xspace}
\newcommand{\surveyansweroptions}[1]{\textit{$\langle$#1$\rangle$}}
\newcommand{\surveyhint}[1]{(Hint: #1)}
\newcommand{\surveyquestion}[1]{\subsubsection{{\textit{\colorbox{aa}{#1}}}}}

\newcommand{\usecase}[1]{#1\xspace}
\newcommand{\new}[1]{\textcolor{black}{#1}}

\newacro{6G}[6G]{sixth-generation technology standard for cellular networks}
\newacro{AS}[AS]{autonomous system}
\newacro{ATI}[ATI]{affinity for technology interaction}
\newacro{BGP}[BGP]{Border Gateway Protocol}
\newacro{BS}[BS]{base station}
\newacro{CDN}[CDN]{content delivery network}
\newacro{CIX}[CIX]{commercial Internet exchange}
\newacro{CSCW}[CSCW]{Computer-Supported Collaborative Work}
\newacro{GDPR}[GDPR]{General Data Protection Regulation}
\newacro{HCI}[HCI]{Human-Computer Interaction}
\newacro{ICT}[ICT]{information and communication technology}
\newacro{iGDB}[iGDB]{Internet Geographic Database}
\newacro{IRB}[IRB]{institutional review board}
\newacro{IoT}[IoT]{Internet of Things}
\newacro{IP}[IP]{Internet Protocol}
\newacro{IQR}[IQR]{inter-quartile range}
\newacro{IRI}[IRI]{instructed response item}
\newacro{ISO}[ISO]{International Organization for Standardization}
\newacro{IXP}[IXP]{Internet exchange point}
\newacro{KDE}[KDE]{kernel density estimation}
\newacro{MNO}[MNO]{mobile network operator}
\newacro{E2EE}[E2EE]{end-to-end encrypted}
\newacro{PoP}[PoP]{point of presence}
\newacro{RAN}[RAN]{radio access network}
\newacro{SRSI}[SRSI]{self-reported single item}
\newacro{UAV}[UAV]{uncrewed aerial vehicle}
\newacro{VoIP}[VoIP]{Voice over Internet Protocol}

\newacro{BW}[BW]{Baden-Württemberg}
\newacro{BY}[BY]{Bavaria}
\newacro{BE}[BE]{Berlin}
\newacro{BB}[BB]{Brandenburg}
\newacro{HB}[HB]{Bremen}
\newacro{HH}[HH]{Hamburg}
\newacro{HE}[HE]{Hesse}
\newacro{MV}[MV]{Mecklenburg-Western Pomerania}
\newacro{NI}[NI]{Lower Saxony}
\newacro{NW}[NW]{North Rhine-Westphalia}
\newacro{RP}[RP]{Rhineland-Palatinate}
\newacro{SL}[SL]{Saarland}
\newacro{SN}[SN]{Saxony}
\newacro{ST}[ST]{Saxony-Anhalt}
\newacro{SH}[SH]{Schleswig-Holstein}
\newacro{TH}[TH]{Thuringia}

\copyrightyear{2025}
\acmYear{2025}
\setcopyright{acmlicensed}
\acmConference[CHI '25]{CHI Conference on Human Factors in Computing Systems}{April 26-May 1, 2025}{Yokohama, Japan}
\acmBooktitle{CHI Conference on Human Factors in Computing Systems (CHI '25), April 26-May 1, 2025, Yokohama, Japan}
\acmDOI{10.1145/3706598.3714324}
\acmISBN{979-8-4007-1394-1/25/04}

\newcommand*{\thetitle}{The User Perspective on Island-Ready 6G Communication: A Survey of Future Smartphone Usage in Crisis-Struck Areas with Local Cellular Connectivity}
\newcommand*{\theshorttitle}{The User Perspective on Island-Ready 6G Communication}
\newcommand*{\theshortauthors}{Janzen et al.}

\begin{document}

\title[\theshorttitle]{\thetitle}

\author{Leon Janzen}
\orcid{0000-0003-2648-6507}
\affiliation{%
    \department{Secure Mobile Networking Lab}
    \institution{TU Darmstadt}
    \city{Darmstadt}
    \country{Germany}
}
\email{ljanzen@seemoo.de}

\author{Florentin Putz}
\orcid{0000-0003-3122-7315}
\affiliation{%
    \department{Secure Mobile Networking Lab}
    \institution{TU Darmstadt}
    \city{Darmstadt}
    \country{Germany}
}
\email{fputz@seemoo.de}

\author{Kolja Straub}
\orcid{0009-0004-4954-6570}
\affiliation{%
\department{Secure Mobile Networking Lab}
\institution{TU Darmstadt}
\country{Germany}
}
\email{kstraub@seemoo.de}

\author{Marc-Andr{\'e} Kaufhold}
\orcid{0000-0002-0387-9597}
\affiliation{%
\department{Science and Technology for Peace and Security}
\institution{TU Darmstadt}
\country{Germany}
}
\email{kaufhold@peasec.tu-darmstadt.de}

\author{Matthias Hollick}
\orcid{0000-0002-9163-5989}
\affiliation{%
\department{Secure Mobile Networking Lab}
\institution{TU Darmstadt}
\country{Germany}
}
\email{mhollick@seemoo.de}

\renewcommand{\shortauthors}{\theshortauthors}

\begin{abstract}
Using smartphone apps during crises is well-established, proving critical for efficient crisis response. However, such apps become futile without an Internet connection, which is a common issue during crises. \new{The ongoing 6G standardization explores the capability to provide local cellular connectivity for areas cut off from the Internet in crises. This paper introduces to the HCI community the concept of cellular \textit{island connectivity} in isolated areas, promising a seamless transition from normal operation to island operation with local-only cellular connectivity}. It presents findings from a survey ($N = 857$) among adult smartphone users from major German cities regarding their smartphone usage preferences in this model. \new{Results show a shift in app demand, with users favoring general-purpose apps over dedicated crisis apps in specific scenarios.} We prioritize smartphone services based on their criticality, distinguishing between apps essential for crisis response and those supporting routines. Our findings provide operators, developers, and authorities insights into making user-centric design decisions for implementing island-ready 6G communication.
\end{abstract}

\begin{teaserfigure}
\centering
\begin{subfigure}[t]{0.275\textwidth}
    \centering
    \includegraphics[width=\linewidth]{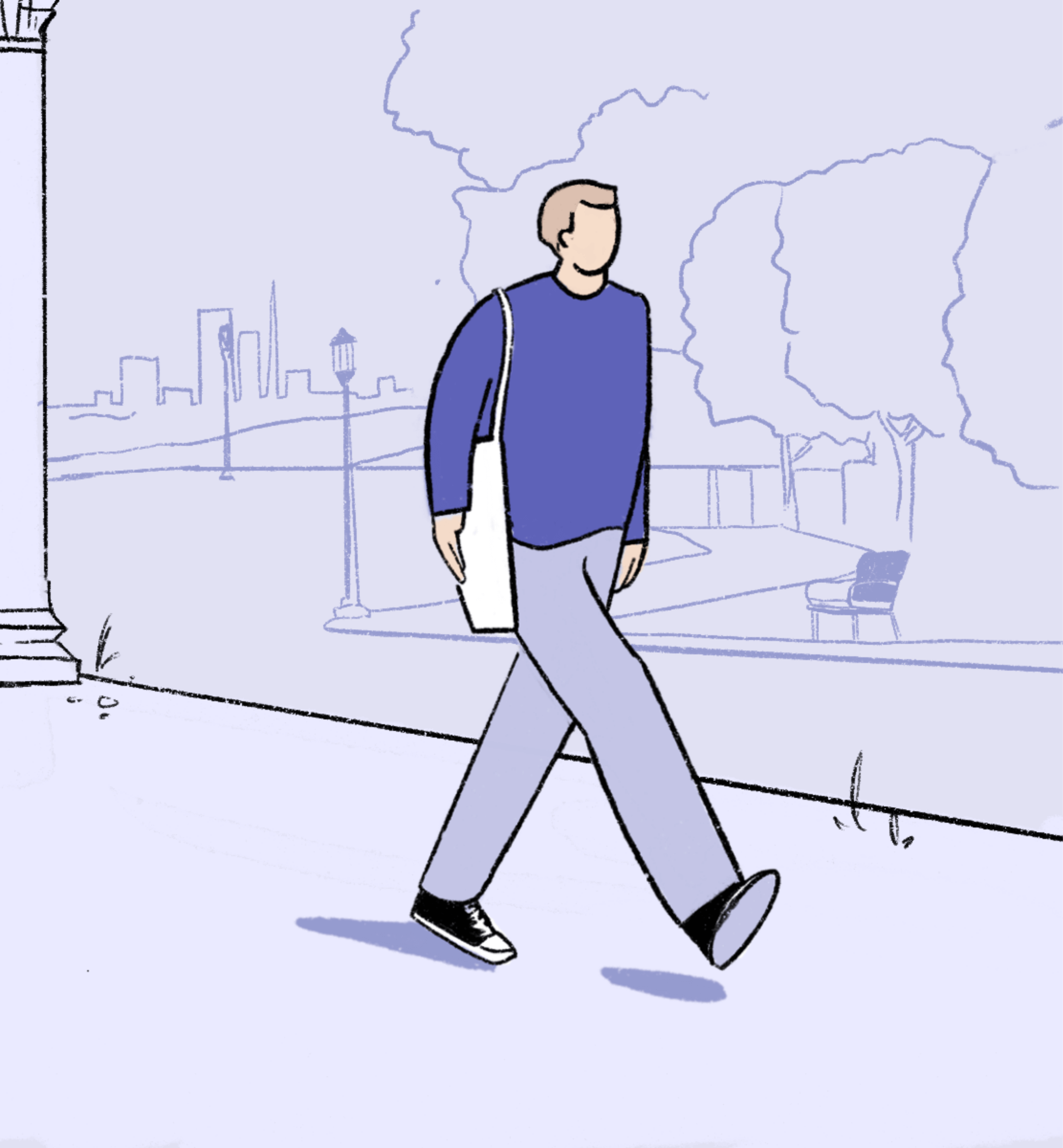}
    \caption{
      You are taking a walk in \colorbox{aa}{\textit{your city.}}
    }
    \label{fig:comic11}
\end{subfigure}
\hspace{\comicmargin}
\begin{subfigure}[t]{0.275\textwidth}
    \centering
    \includegraphics[width=\linewidth]{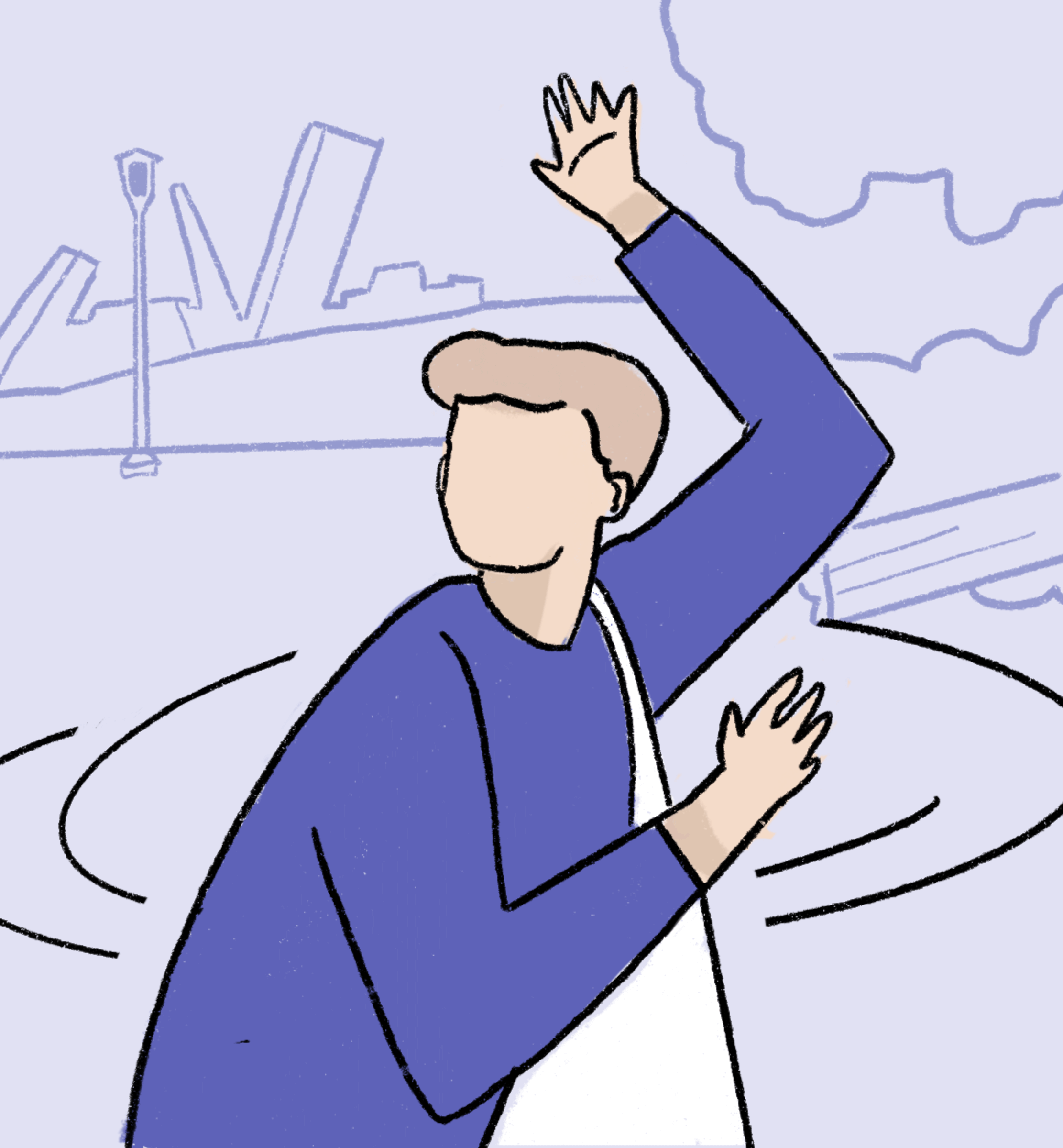}
    \caption{
      Suddenly, you feel several strong tremors and notice that everything around you is shaking.
    }
    \label{fig:comic12}
\end{subfigure}
\hspace{\comicmargin}
\begin{subfigure}[t]{0.275\textwidth}
    \centering
    \includegraphics[width=\linewidth]{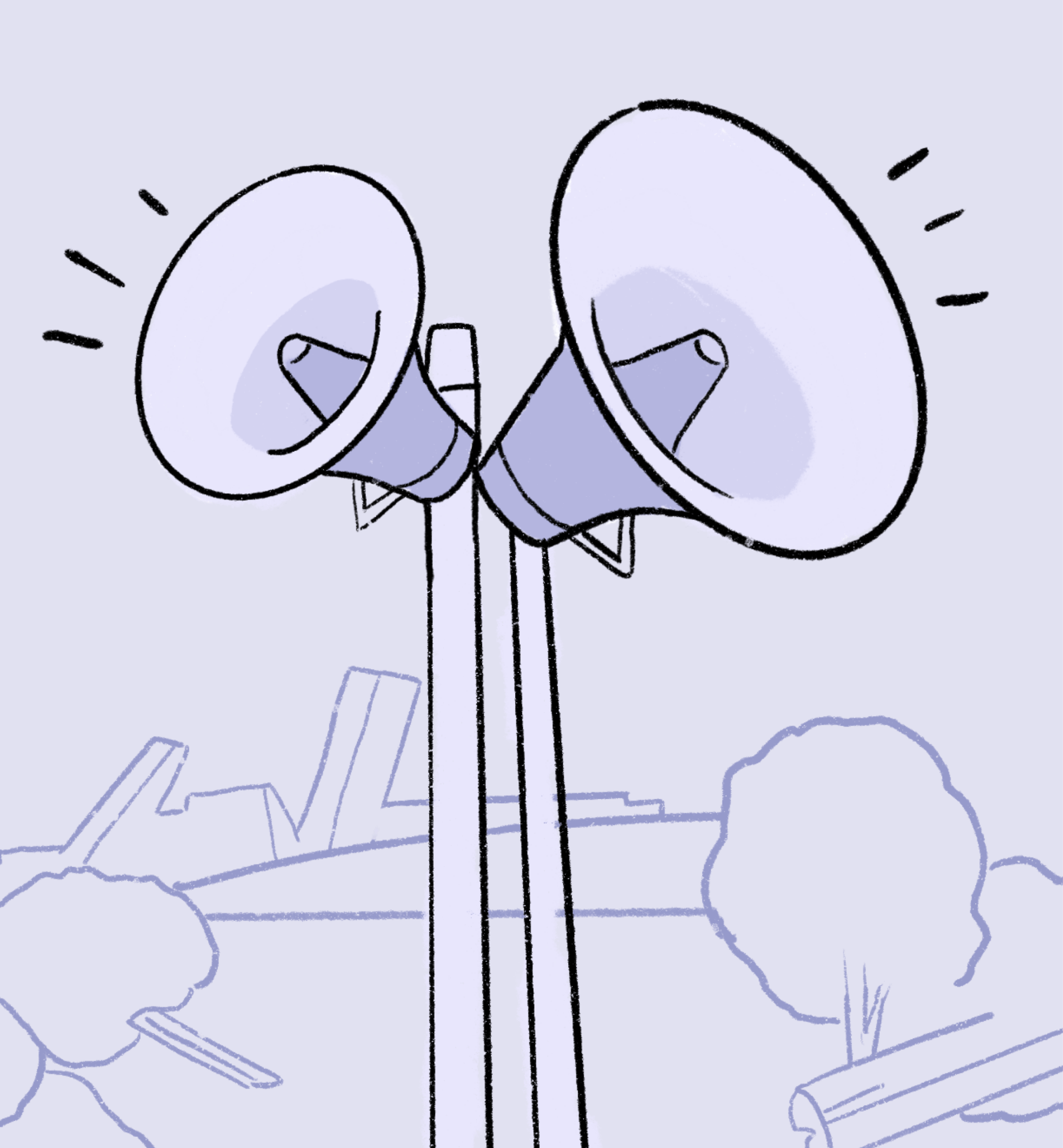}
    \caption{
      A siren sounds with a loudspeaker announcement: "Attention! The city has been hit by an earthquake. Please seek safety."
    }
    \label{fig:comic13}
\end{subfigure}
\caption{
    \captionheadline{Crisis Scenario}
    This comic board introduced our participants to the assumed crisis scenario of our study, an earthquake.
    We individualized the first comic (Figure~\ref{fig:comic11}) by replacing \colorbox{aa}{\textit{your city}} with the participant's city.
}
\label{fig:comic1}
\Description{
    A comic board with three side-by-side comics.
    In Comic B, a person walks in a park.
    They look relaxed with their hands in the pockets of their trousers and they are carrying a bag on their right shoulder.
    There is a calm park in the background.
    In Comic B, we see a close-up of the person with their hands in the air as if trying not to fall over.
    In the background, the park is shaking, and a tree has fallen over.
    Comic C shows a close-up of a loudspeaker mounted on a pole in the park.
    The loudspeaker looks like it is making an announcement.
    In the background, we see another angle of the park with multiple fallen-over trees.
}
\end{teaserfigure}



\maketitle

\section{Introduction}
\label{sec: introduction}
\acresetall

For the past 20 years, the use of social media during emergencies and natural disasters has dramatically increased \cite{reuter_social_2018}, thus receiving substantial attention within the research fields of \ac{HCI} and \ac{CSCW}.
As more and more people own a smartphone and carry it with them at all times \new{\cite{bundesamt_destatis_ausstattung_2022,statista_2024_number}}, smartphones have become a central tool for users to cope with crisis scenarios.
\new{While social media and dedicated crisis apps have been well-researched in the community, e.g., by \cite{eismann_collective_2016,tan_mobile_2017}, the role of other general-purpose smartphone apps} in crisis response has not received much attention \cite{simko_use_2023}.
To motivate further efforts in this field, our first research question is:

\aptLtoX{\begin{aptdispbox}
\researchquestion{RQ1:} Which apps do citizens prefer to use for crisis-specific use cases?
\end{aptdispbox}}{\begin{shaded*}
  \researchquestion{RQ1:} Which apps do citizens prefer to use for crisis-specific use cases?
\end{shaded*}}

One area of concern for smartphone apps during crisis response is that Internet outages often accompany crises, but many smartphone apps require an Internet connection for full functionality.
\new{For example, the resilience strategy of the German federal network agency considers natural disasters, large-scale cyber attacks, and destruction during armed conflicts as threat scenarios for telecommunications networks \cite{abteilung_2_resilience_2022}.}
In such scenarios where crisis-struck areas are also affected by Internet outages, users cannot use smartphone apps that require Internet connectivity.
Furthermore, cellular services, such as emergency calls, telephony, and cell broadcasts, are unavailable as they require connection to the cellular core network that is most likely hosted outside the affected area.

With the upcoming standardization of the \ac{6G}, there are proposals to distribute the core network, primarily to achieve low latencies \cite{mukherjee_distributed_2018,li_joint_2021,chen_joint_2024,corici_shortcuts_2024}.
However, decentralizing \ac{6G} also increases the probability that crisis-struck areas have a core network in proximity, which operators could use to maintain local cellular Internet connectivity within the affected area.
Section~\ref{sec: background} defines such areas as \textit{islands,} as they form a self-contained network isolated from the outside Internet, and introduces the conceptual model of \textit{island connectivity}.
We envision users on islands using their smartphone apps locally, giving them access to crisis-relevant apps and letting them communicate with friends on the island.
Other apps and services, however, would not be available on islands, e.g., smartphone apps that require remote server connections and telephony with people outside the island.
Considering the findings of our narrative review \cite{pare_synthesizing_2015}, understanding the user-smartphone interaction on islands with local cellular connectivity is an unstudied research problem.
We contribute to this field with our second research question:

\aptLtoX{\begin{aptdispbox}
\researchquestion{RQ2:} Which apps do citizens prefer to use in isolated areas with local connectivity?
\end{aptdispbox}}{\begin{shaded*}
\researchquestion{RQ2:} Which apps do citizens prefer to use in isolated areas with local connectivity?
\end{shaded*}}

We designed and conducted a survey ($ N = \nfinal $) to understand smartphone usage with island connectivity and answer research questions \researchquestion{RQ1} and \researchquestion{RQ2}.
Section~\ref{sec: method} discusses our study design and sample representative of general adult smartphone users in major German cities.
Our findings indicate that there is demand for general-purpose apps in crisis response, e.g., users prefer messengers and news apps for communication and information.
We studied which apps users prefer with island connectivity, finding a demand increase for some apps, e.g., messengers, news, and crisis apps, and a demand decrease for others, e.g., sports, shopping, and travel apps.
Section~\ref{sec: results} describes our findings in detail.

Putting the vision of \textit{island connectivity} into practice poses many challenges and involves many stakeholders:
Storage and compute capabilities must be available in a distributed manner, app developers must adapt their backend communication infrastructure to equip smartphone apps for island operation, and \acp{MNO} must prepare \ac{6G} to transition into island operation when a crisis-struck area is disconnected from the Internet.
These efforts must be either justified by user demand or regulated by authorities.
For instance, the resilience strategy of the German federal network agency requests that "citizens should be able to use a certain range of basic services even in the event of widespread [outages]" \cite{abteilung_2_resilience_2022}.
While they name emergency calls and cell broadcasts, they do not regulate how to prioritize apps when faced with limited capacity.
Our third research question addresses this gap:

\aptLtoX{\begin{aptdispbox}
  \researchquestion{RQ3:} How should operators prioritize smartphone services in isolated areas with local connectivity?
\end{aptdispbox}}{\begin{shaded*}
    \researchquestion{RQ3:} How should operators prioritize smartphone services in isolated areas with local connectivity?
\end{shaded*}}

We construct a \textit{smartphone service typology for island connectivity,} i.e., systematically classifying smartphone apps and cellular services according to their criticality for crisis response on islands.
Our typology, discussed in detail in Section~\ref{sec: discussion}, distinguishes smartphone services that are crucial for crisis response, e.g., making emergency calls, receiving cell broadcasts, and using messengers, and those that enable crisis routines, e.g., finance apps, social media, and telephony.
We compile our survey results and the typology into design implications that support stakeholders in making user-centric design decisions while realizing island-ready 6G communication.

\section{Related Work and Research Gap}
\label{sec: related-work}

This section reviews related user studies and highlights research gaps in prior work that we address in our work.

\subsection{\new{App Usage During Crises}}

The first decade of crisis informatics research had a strong emphasis on social media, providing qualitative empirical case studies \cite{olteanu_what_2015}, identifying usage and role patterns \cite{starbird__2011,reuter_combining_2013}, designing algorithms for processing big crisis data \cite{castillo_big_2016,imran_processing_2015}, and evaluating open source and social media analytics systems for enhanced situational awareness \cite{alam_descriptive_2020,kaufhold_mitigating_2020,kaufhold_we_2024}.
Many quantitative representative surveys examined citizens' usage patterns, perceived usefulness, and expectations towards social media in crises \cite{reuter_increasing_2023} and compared perceptions across European countries, North America, and China \cite{reuter_impact_2019,wang_understanding_2022}.

However, considering the narrow focus of crisis informatics, researchers have suggested looking beyond social media \cite{soden_informating_2018}.
While general-purpose apps and social media sometimes include functionality for community-oriented crisis response \cite{tan_mobile_2017}, multiple official so-called crisis and warning apps have been established across various countries of the Global North \cite{hauri_comparative_2022}.
These often provide location-based warnings and tips on how to behave before, during, and after an emergency, and sometimes facilitate communication with emergency services \cite{dallo_why_2021}.
Existing qualitative representative surveys with German citizens show increasing adoption of crisis apps, highlighting their perceived usefulness and the importance of different functionalities \cite{haunschild_perceptions_2022}.
Researchers have examined the motivation and usability factors for using crisis apps \cite{fischer-presler_protection-motivation_2022,tan_usability_2020}, including nudging approaches to enhance preparedness \cite{haunschild_preparedness_2023}.
However, considering the findings of our narrative review \cite{pare_synthesizing_2015}, there is a lack of research examining the citizens' preferences for different apps for crisis-specific use cases.
We contribute to this area by studying which apps citizens prefer to use for crisis-specific use cases and testing the demand for non-crisis apps during crises (\researchquestion{RQ1}).

\subsection{\new{Communication Networks During Crises}}

Existing studies on social media and crisis apps often lack attention to the underlying infrastructure requirements, such as energy and telecommunications, which are especially important in isolated areas.
Not least, the 2021 Ahr Valley flooding showcased the sociotechnical complexity of warnings in a developed country like Germany.
On the one hand, the acute warning period was too short; on the other hand, communication and electricity infrastructure failed during the floods and thus impaired further crisis communication and rescue operations \cite{fekete_here_2021}.
Data from an online survey with citizens reveals that \SI{35}{\percent} of the respondents from the federal state of North Rhine-Westphalia and \SI{29}{\percent} from Rhineland-Palatinate did not receive any warning, highlighting the importance of effective administrative decision-making processes and resilient warning systems.
Of those warned, \SI{85}{\percent} did not expect severe flooding, and \SI{46}{\percent} reported a lack of situational knowledge on protective behavior \cite{thieken_performance_2023}.
Another representative survey revealed that German citizens demand behavioral instructions (\SI{83}{\percent}), information on the reason for the failure (\SI{80}{\percent}), and the expected outage duration (\SI{78}{\percent}) from the respective infrastructure provider \cite{kaufhold_potentiale_2019}.

\subsection{\new{Related Notions of Island Connectivity}}
The idea of \emph{islands} has a long history in networking and disaster research \cite{calegari_1997_parallel,palmier_2013_analyzing}.
Our interpretation of an island is that a region affected by a crisis and simultaneously disconnected from the outside Internet can maintain local Internet connectivity for users, e.g., to app servers hosted on the network edge, allowing users to continue using general-purpose apps on their smartphones.
Section~\ref{sec: background} elaborates on the conceptual model of island connectivity.
In the following, we distinguish island connectivity from related notions.

A central problem of modern ICT for crisis communication is their dependence on centralized infrastructures \cite{kaufhold_exploring_2024}.
Thus, it is essential to harden centralized infrastructures against natural hazards \cite{kuntke_rural_2023} and explore decentralized crisis communication.
Related proposals exist to enhance the resilience of cellular networks, e.g., with self-organizing small cell networks \cite{zhang_2016_self} or network-in-a-box approaches \cite{pozza_2018_network}.
These approaches are often limited to a specific part of the cellular network or require time to set up during a crisis.
In contrast, island connectivity considers the entire 6G system, and we envision that island-ready 6G can seamlessly switch to island operation in case of a crisis.
While related work studied decentralized and disruption-tolerant networks in smart rural areas \cite{hochst_mobile_2023,kuntke_how_2023}, the prevalence of \ac{IoT} devices will be exceptionally high in smart cities, including \acp{UAV}, smart cars, smart lanterns, and smartphones, for the establishment of multi-hop networks \cite{sterz_energy-efficient_2023,heise_optimized_2022}.
Since applications and services will need to switch between different modes of centralized and decentralized operation, probably with varying functionalities available, HCI has to ensure a seamless user experience during these transitions \cite{haesler_connected_2021}.
Yet, we did not find any study of citizens' preferences regarding different app categories in isolated areas embracing the potential of island connectivity.
To fill this gap, we investigate which apps users prefer to use in crisis-struck areas isolated from the Internet (\researchquestion{RQ2}).

\subsection{\new{Practical Aspects of Island Connectivity}}
Putting the vision of \textit{island connectivity} into practice poses many challenges and involves many stakeholders.
App developers must adapt their backend communication infrastructure to ready smartphone apps for island operation.
\Acp{MNO} must prepare \ac{6G} to transition into island operation when a crisis-struck area is disconnected from the Internet.
These efforts must be either justified by user demand or regulated by authorities.
To our knowledge, the user demand for islands is an unstudied research problem.
Thus, the research community lacks an understanding of which app categories citizens prefer to use in isolated areas.
Furthermore, authorities should regulate how \acp{MNO} prioritize smartphone apps when faced with limited capacity.
However, no existing regulation guides \acp{MNO} in this endeavor \cite{abteilung_2_resilience_2022}.
We address this gap by constructing a smartphone service typology (\researchquestion{RQ3}).

\new{In summary, \textbf{our study contributes to understanding citizens' smartphone usage in crisis-struck isolated areas,} motivating the demand for general-purpose apps for crisis-specific use cases (\researchquestion{RQ1}), reporting the demand for smartphone apps in isolated areas with local connectivity (\researchquestion{RQ2}), and suggesting prioritization of smartphone services for these scenarios (\researchquestion{RQ3}).}
\Cref{ssec: background-related-concepts} complements our review of prior work by distinguishing our \textit{island connectivity} model from related concepts.

\section{The Conceptual Model of Island Connectivity}
\label{sec: background}

\new{In this section, we introduce the conceptual model of \textit{island connectivity}, which is our study's specific 
application area.}

\begin{figure}
    \centering
    \includegraphics[width=\columnwidth]{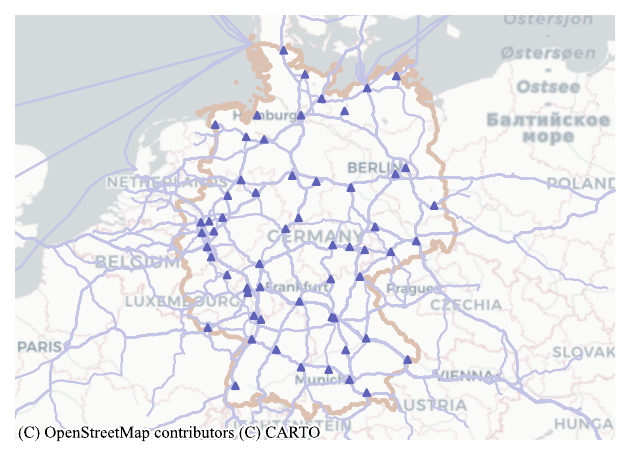}
    \caption{
        \captionheadline{Internet Infrastructure in Germany}
        \textcolor{mybeige}{\textbf{A beige line}} shows the German border,
        \textcolor{myblue}{\textbf{blue triangles}} highlight known physical interconnection facility locations, e.g., \acfp{PoP}, \acfp{IXP}, or cable landing sites, and 
        \textcolor{mypurple}{\textbf{purple lines}} show the routes of submarine cables and the assumed terrestrial  cable routes, following roadways, rails, and power lines \cite{anderson_igdb_2022,durairajan_intertubes_2015}.
    }
    \label{fig:german-cities}
    \Description{
        The background shows a gray-on-white map of Europe with the German border highlighted in a beige line.
        Germany is located at the figure's center, with the German-Danish border and the North and East Sea in the upper center and the German border to Switzerland and Austria in the lower center.
        A part of the United Kingdom is visible on the center-left edge.
        To the right, the map shows the entirety of Czechia and ends west of the Polish city of Krakow.
        There are many blue triangles positioned across Germany, mostly co-located with major cities.
        There are many lines printed in light blue that connect the triangles but also cross the German border in all directions.
        Cables are also traversing the North Sea and East Sea in the upper part of the figure.
    }
\end{figure}

\subsection{Background: Internet Connectivity}
\label{ssec: background-internet-connectivity}

The Internet is a network of networks that builds a physical and logical graph around the globe \cite{braden_rfc1122_1989}.
The physical Internet comprises infrastructure nodes, e.g., \acfp{IXP}, colocation centers, and \acfp{PoP}, connected by fiber conduits, e.g., terrestrial long-haul and submarine cables \cite{anderson_igdb_2022}.
The \ac{IP} suite defines abstraction layers to enable logical end-to-end connections while abstracting from the concrete physical connections in use, e.g., allowing end-to-end communication between hosts based on \ac{IP} addresses \cite{mccauley_extracting_2023}.

When users open a smartphone app that requires Internet connectivity to connect to an app server, the app usually stores the server's domain name or \ac{IP} address.
The smartphone first connects to a 4G/5G base station to establish a connection to the app server, from where the user data is augmented with control plane data and sent to the \ac{MNO}'s core network \cite{project_3gpp_system_2023}.
If the core network successfully processes the control plane data, e.g., handling authentication and session management, the user data proceeds to the Internet.
Alternatively, smartphones can connect to the Internet via a Wi-Fi router.
On the Internet, the user data is routed toward the app server's \ac{IP} address until it reaches its destination.

Current cellular networks have a mostly central architecture with limited core network gateways per country, which can cause path inflation \cite{zarifis_diagnosing_2014}, i.e., a diversion from the shortest route due to the requirement of passing the core network.
Path inflation is especially pronounced if the app server is hosted on the local network edge close to the user but the closest core network gateway is remote.
The physical Internet backbone in Germany is considered critical infrastructure and, therefore, is not publicly available \cite{bundesnetzagentur_informationsplattform_2024}.
However, Figure~\ref{fig:german-cities} visualizes how the \ac{iGDB} \cite{anderson_igdb_2022} interpolates connections between known Internet infrastructure, assuming that cables are often installed along existing networks such as roadways, rail, and power lines \cite{durairajan_intertubes_2015}.

\subsection{Islands: Crisis-Struck Areas Isolated From the Internet}

The communication infrastructure is generally expected to be resilient against outages.
Still, cellular network outages are actively considered in the resilience strategy of the German federal network agency \cite{abteilung_2_resilience_2022}, and previous outages have isolated individual areas \cite{mastd_mobilfunk_2021}.
Another complication is that user traffic has to traverse the core network before being forwarded to the Internet, but the chances of a cellular core network being available in the affected area are negligible due to the low geo-redundancy of cellular core network gateways \cite{zarifis_diagnosing_2014}.
These gateway locations of the German \acp{MNO}, i.e., \textit{1\&1 AG,} \textit{Deutsche Telekom AG,} \textit{Telef{\'o}nica Deutschland Holding AG,} \textit{Vodafone GmbH,} are unknown.
Still, observations from other countries \cite{rula_adopting_2017} suggest only a handful of gateways per \ac{MNO} exist in Germany.
As a result, users in affected areas likely cannot use 4G/5G connectivity to reach these servers despite the locally intact physical Internet and servers being available on the local network edge.
We aggregate the notion of a crisis-struck area that is isolated from the outside Internet in the definition of \textit{islands}:

\aptLtoX{\begin{aptdispbox}
   An \textbf{island} is a crisis-struck area isolated from the outside Internet, preventing citizens from using smartphone apps that require an Internet connection to remote servers.
\end{aptdispbox}}{\begin{shaded*}
    An \textbf{island} is a crisis-struck area isolated from the outside Internet, preventing citizens from using smartphone apps that require an Internet connection to remote servers.
\end{shaded*}}

Several terms exist in the literature to describe deviations from regular operation.
This paper uses the term \textit{crisis} as a hypernym for crises, emergencies, catastrophes, and disasters \cite{edwards_crisis_2022}.
Islands can be of various shapes and sizes, ranging from small rural areas to major cities.
One example is the 2021 flooding in Western Europe, where the floods destroyed fiber links installed under roads and bridges and, thus, disconnected large parts of the Ahr Valley from the Internet for days to weeks \cite{mastd_mobilfunk_2021}.
While satellite connectivity has successfully been used in crisis-struck areas to provide citizens with Internet access, the achievable bandwidth with a few satellite links is insufficient to reconnect an entire region to the Internet \cite{michel_first_2022}.
Another common approach is microwave transmissions, closing the gap of broken fiber links with point-to-point directional antennas \cite{deepak_overview_2019}.
However, their applicability depends on the type of failure responsible for the isolation, and setting up directional antennas can take several days.

\subsection{Island Connectivity: Local Fall-Back Cellular Networks}

With multi-access edge computing, operators increasingly aim to move content and computational power closer to the user by hosting servers on the network edge, e.g., co-located with \acp{IXP} or in a data center within the user's proximity \cite{bahrami_edge_2023}.
These advances reduce latency as Internet traffic destines within the user's proximity instead of being routed to a remote data center.
Still, they do not fully apply to cellular network users due to \textit{path inflation} \cite{zarifis_diagnosing_2014,zhang_inferring_2021,kiess_centralized_2014}.
With \ac{6G} on the horizon, there are proposals \cite{mukherjee_distributed_2018,li_joint_2021,chen_joint_2024,corici_shortcuts_2024} to tackle the bottleneck of path inflation by distributing core network functions, e.g., co-located at suitable Internet nodes or generally within the data centers of major cities.

A higher geo-redundancy of core network gateways would improve latencies and the chances of \textit{islands} having a core network available.
As a result, island users could use crisis-relevant cellular services, i.e., making emergency calls, receiving cell broadcasts, and contacting close ones on the island by text or phone call.
In addition, a core network on the island facilitates cellular Internet connectivity, thus enabling users to use smartphone apps for crisis response.
In this context, app servers must be deployed within the island's boundaries, aligning with the multi-access edge computing trend.
To this end, we distinguish \textit{local} and \textit{remote} connections, the former staying within the island's bounds, while the latter destine at a server outside the island.
We distinguish two operation modes of the cellular network: \textit{normal operation} in non-crisis times and \textit{island operation,} providing island connectivity on islands.

\aptLtoX{\begin{aptdispbox}
  \textbf{Island connectivity} enables users on an island to use the cellular network within the island's bounds, facilitating the use of cellular services (e.g., emergency calls and cell broadcasts) within the island and access to servers on the islands's network edge (e.g., multi-access edge computing).
\end{aptdispbox}}{\begin{shaded*}
    \textbf{Island connectivity} enables users on an island to use the cellular network within the island's bounds, facilitating the use of cellular services (e.g., emergency calls and cell broadcasts) within the island and access to servers on the islands's network edge (e.g., multi-access edge computing).
\end{shaded*}}

With island connectivity, users can access the cellular network within that island's boundaries, i.e., they can use apps with app servers hosted on the local edge but not apps with remote app servers; they can communicate with friends on the island but not with friends outside the island.
\new{One requirement of island connectivity is an intact power supply, another critical infrastructure that can fail during natural disasters or be affected by dedicated cyber attacks.}
The German federal network agency reports that many \acp{MNO} already have emergency and backup power systems.
Still, these are often severely limited, providing only a few hours of electricity.
For the future, it recommends considering renewable energies as on-site backups and for end users to have a mobile power supply in stock to recharge mobile devices in crisis scenarios \cite{abteilung_2_resilience_2022}.
Implementing the recommendations from these guidelines ensures that island users can use island connectivity when available \cite{project_3gpp_system_2023}.

The realization of island connectivity poses various challenges to operators, developers, and authorities.
\new{To our knowledge, the HCI community has not studied the user-smartphone interaction in the specific application area of islands with local cellular connectivity.}
This paper aims to understand citizens' smartphone usage on islands and derive design criteria for apps and 6G to facilitate user-centric design decisions while realizing island connectivity.

\subsection{Distinction from Related Technical Concepts}
\label{ssec: background-related-concepts}

The design principles underlying the Internet are simple and mostly unchanged since its inception \cite{mccauley_extracting_2023}, as the principles of routing, reliability, and resolution allowed the Internet to grow from a few computers to today’s global scale.
Originally, the Internet was island-ready due to the mostly static configurations and its decentralization, but over time, lost its island capabilities since essential functions, e.g., domain name resolution, were centralized.
Some technical concepts propose to re-integrate decentralized capabilities into the Internet, e.g., peer-to-peer systems as application layer overlays \cite{ripeanu_peer_2001}, \acp{CDN} \cite{pathan_content_2008}, the decentralized web \cite{raman_challenges_2019}, and campus networks \cite{rischke_campus_2021}.
Out of these concepts, only campus networks are related to the cellular network.
Campus networks are private wireless communication networks connecting the users of an organization in a dedicated geographical area, e.g., a university or an industrial site \cite{corici_organic_2023}.
The main objective of 5G campus networks is to use the advantages of cellular connectivity, i.e., higher speed and lower latency.
Campus networks are designed for static settings, i.e., the operator registered the campus network with the authorities and got permission to operate the campus network in a specified frequency band in a specified geographic area.
In addition, only a pre-defined set of SIM cards can connect to the campus network.
As such, the concept of campus networks is clearly distinct from island connectivity by two factors:
First, island connectivity applies to the \textit{public} cellular network and should be usable for all users.
Second, island-ready 6G communication can \textit{dynamically} switch from normal operation to island operation when a core network replica detects a disconnection from the outside Internet.
\new{Nevertheless, non-public networks can benefit from the advances of island connectivity, e.g., when multiple campus networks hosted at different sites that are usually connected through the Internet become isolated due to outages.}

\subsection{Relevance of Island Scenarios}
\label{ssec: background-relevance}

The limited period of island scenarios, infrastructure requirements, and recent proposals for alternative technologies to recover connectivity provoke the question of how relevant island connectivity is for the research community.
First, \ac{ICT} is a critical infrastructure \cite{bundesamt_fur_bevolkerungsschutz_und_katastrophenhilfe_informationstechnik_2024,cybersecurity_and_infrastructure_security_agency_information_2024,cybersecurity_and_infrastructure_security_agency_communications_2024}, underscoring the crucial value of connectivity for crisis response.
In particular, the cellular network plays a pivotal role in crisis response, providing Internet connectivity, emergency calls, and cell broadcasts.

We argue that \textbf{island connectivity can have a critical impact during the first few days of a crisis or even across multiple months in the case of a large-scale conflict,} such as in the ongoing Russo-Ukrainian War, where critical infrastructures are operational targets by cyber or physical attacks \cite{singla_analysis_2023,zarembo_smartphone_2024}.
Past crises underscore the scenario's plausibility and that alternative recovery means do not suffice to reconnect entire areas.
Understanding how citizens prefer to use their smartphones on islands is crucial to supporting the ongoing development of \ac{6G}.
Operators, developers, and authorities face many challenges in implementing this vision, and contributions from the \ac{HCI} community can help facilitate user-centric design decisions.

\section{Method}
\label{sec: method}

Six researchers and one designer were involved in the design phase of this study, cooperating in multiple workshops to combine knowledge from the fields of resilient cellular networks, \ac{HCI}, and \ac{CSCW}.

\subsection{Material}

Our goal was to study the user-smartphone interaction in island connectivity.
We realized early in the design phase of our study that the novelty and complexity of island connectivity as a conceptual model might be hard to grasp.
To avoid overloading participants, we decided to fix a concrete example for islands in our study.
Thus, \textbf{our survey investigates smartphone usage in major German cities,} simplifying the abstract concept of island connectivity to the concrete instantiation of local connectivity in an isolated urban area.
We motivate this choice by the increasing share of the worldwide population living in urban areas, as \SI{71}{\percent} of the German population lived in metropolitan areas in 2022 \cite{bundesamt_destatis_grosstadtregionen_2024}.
In addition, cities play a central role in the Internet, and most of the physical Internet infrastructure aligns around cities \cite{anderson_igdb_2022}.
As such, cities are prime candidates to host replicas of the distributed \ac{6G} core network, which aligns with the concept of resilient cities \cite{zhu_smart_2020,hollick_emergency_2019,khaloopour_resilience-by-design_2024}.

\setlength{\comicwidth}{0.24\textwidth}
\addtolength{\comicwidth}{-0.5pc}
\setlength{\comicmargin}{0.01\textwidth}

\aptLtoX{\begin{figure}
    \centering
    \includegraphics[width=\columnwidth]{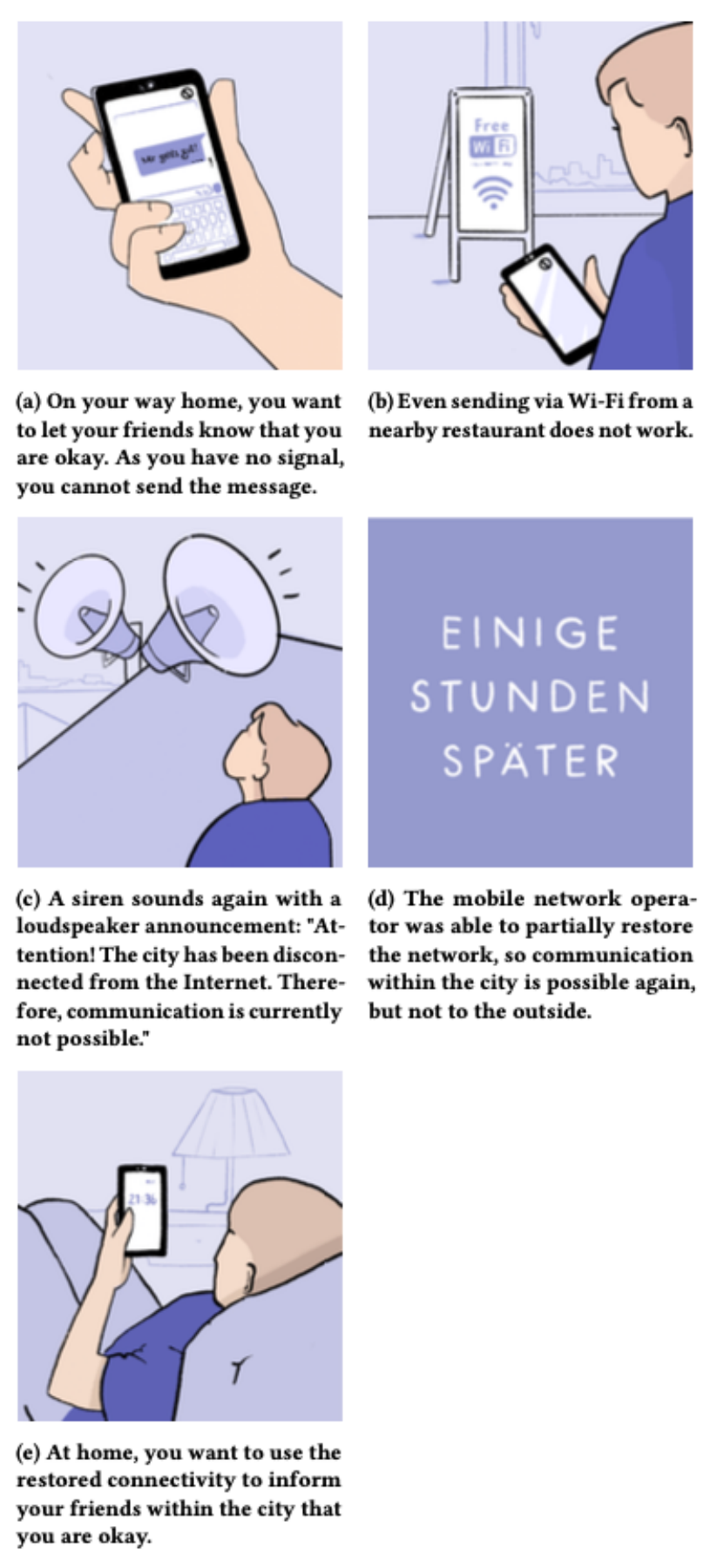}
  \caption{
        \captionheadline{Local Connectivity Model}
        This comic board introduced our participants to the conceptual model of local connectivity in isolated cities.
        The message in Figure~\ref{fig:comic21} translates to "I'm fine!".
        The text in Figure~\ref{fig:comic24} translates to "A few hours later."}
   \label{fig:comic24}
  \label{fig:comic21}
    \label{fig:comic2}
\end{figure}}{\begin{figure}
    \centering
    
    \begin{subfigure}[t]{\comicwidth} 
        \centering
        \includegraphics[width=\linewidth]{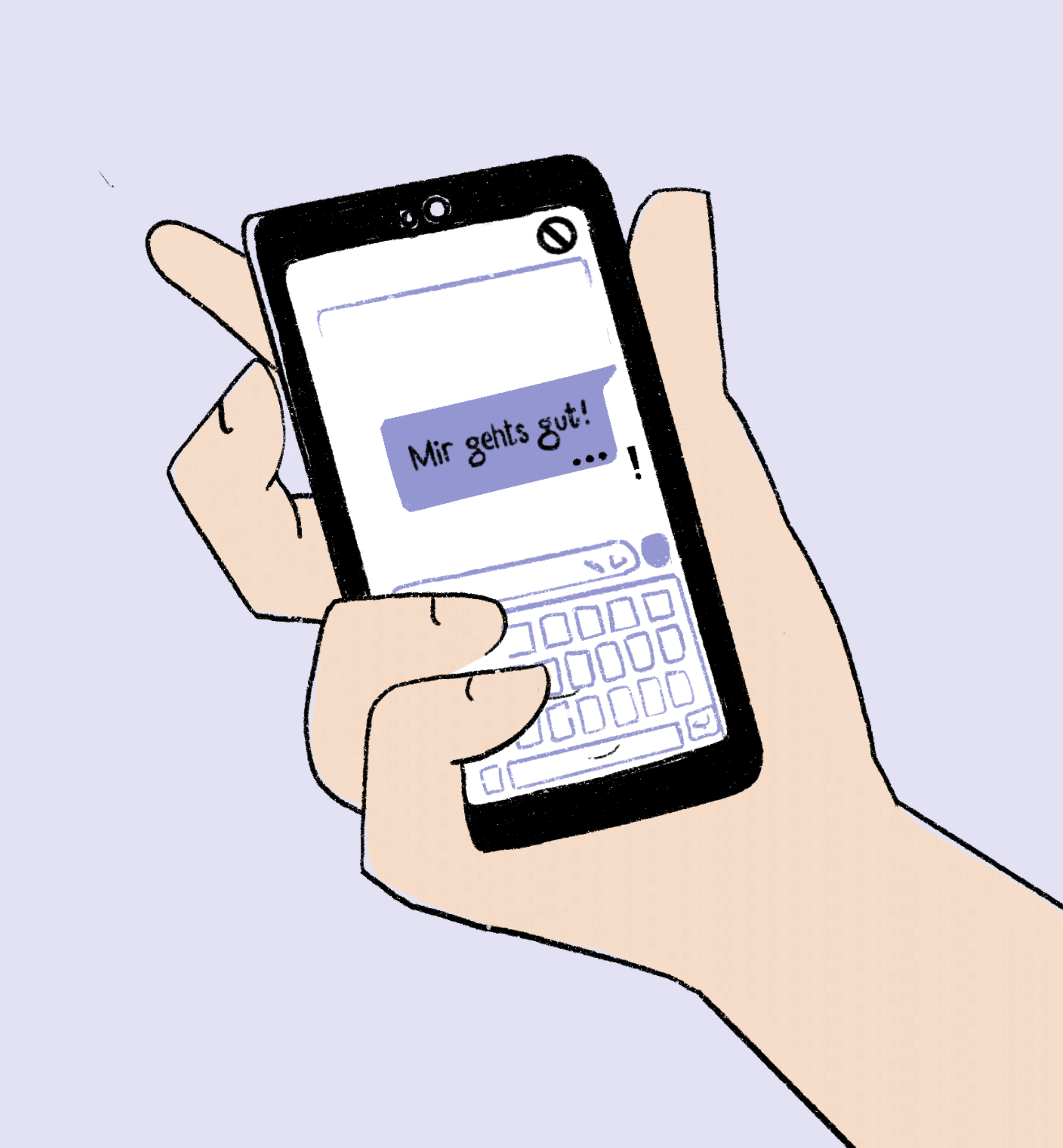}
        \caption{
            On your way home, you want to let your friends know that you are okay.
            As you have no signal, you cannot send the message.
        }
        \label{fig:comic21}
    \end{subfigure} 
    \hspace{\comicmargin} 
    \vspace{\comicmargin} 
    \begin{subfigure}[t]{\comicwidth} 
        \centering
        \includegraphics[width=\linewidth]{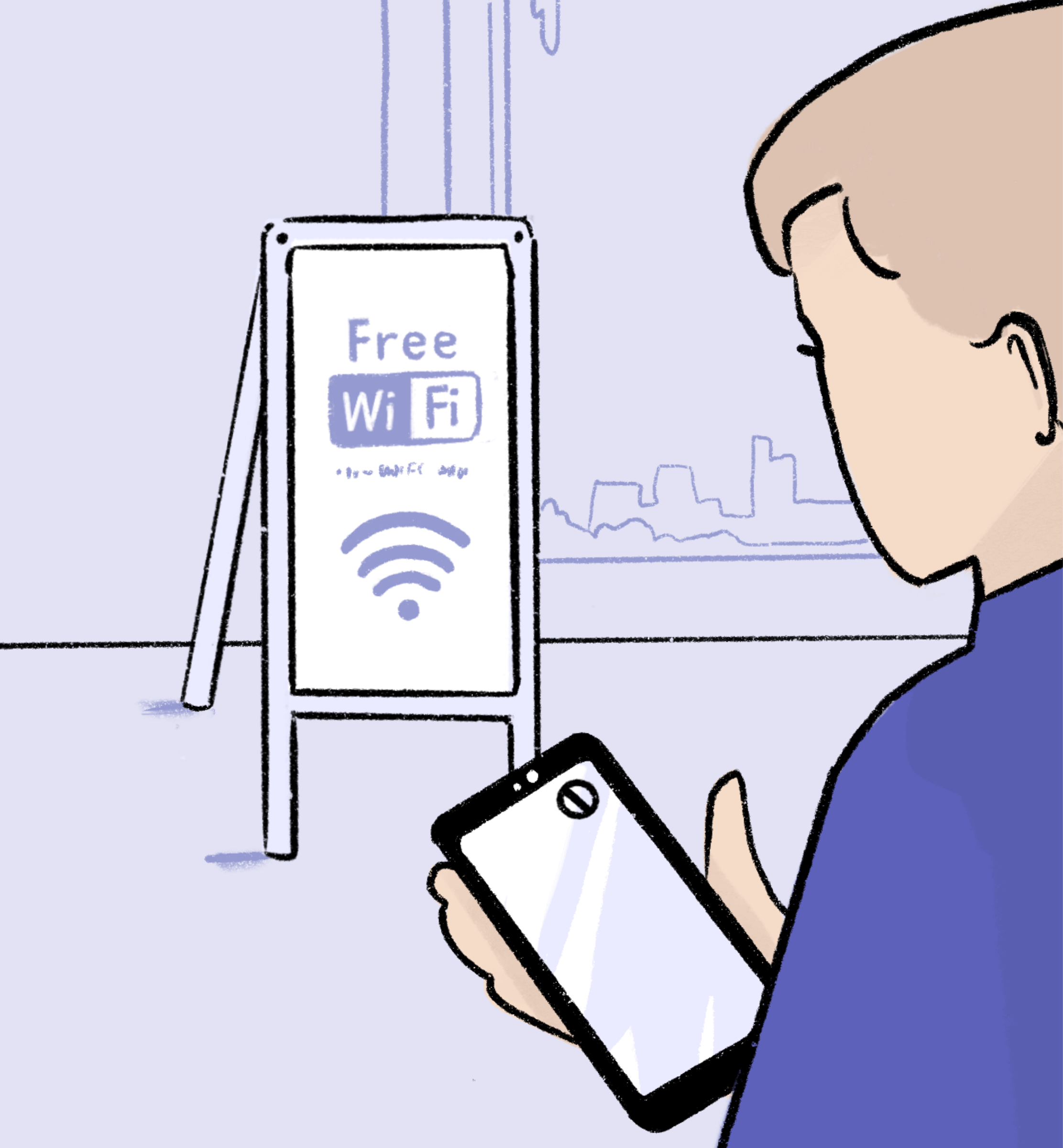}
        \caption{
            Even sending via Wi-Fi from a nearby restaurant does not work.
        }
        \label{fig:comic22}
    \end{subfigure} 
    \begin{subfigure}[t]{\comicwidth} 
        \centering
        \includegraphics[width=\linewidth]{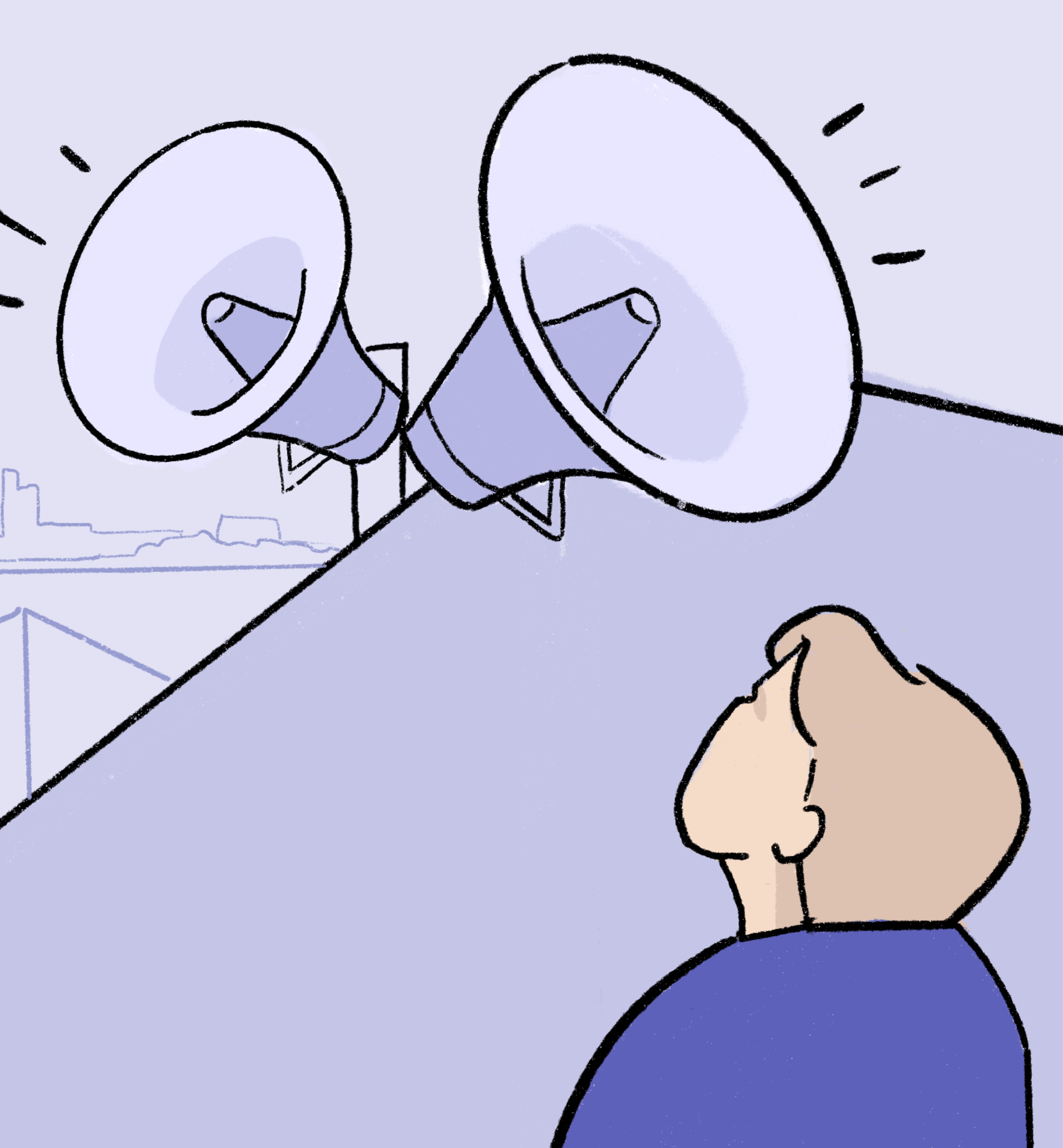}
        \caption{
            A siren sounds again with a loudspeaker announcement:
            "Attention! The city has been disconnected from the Internet.
            Therefore, communication is currently not possible."
        }
        \label{fig:comic23}
    \end{subfigure} 
    \hspace{\comicmargin} 
    \vspace{\comicmargin} 
    \begin{subfigure}[t]{\comicwidth} 
        \centering
        \includegraphics[width=\linewidth]{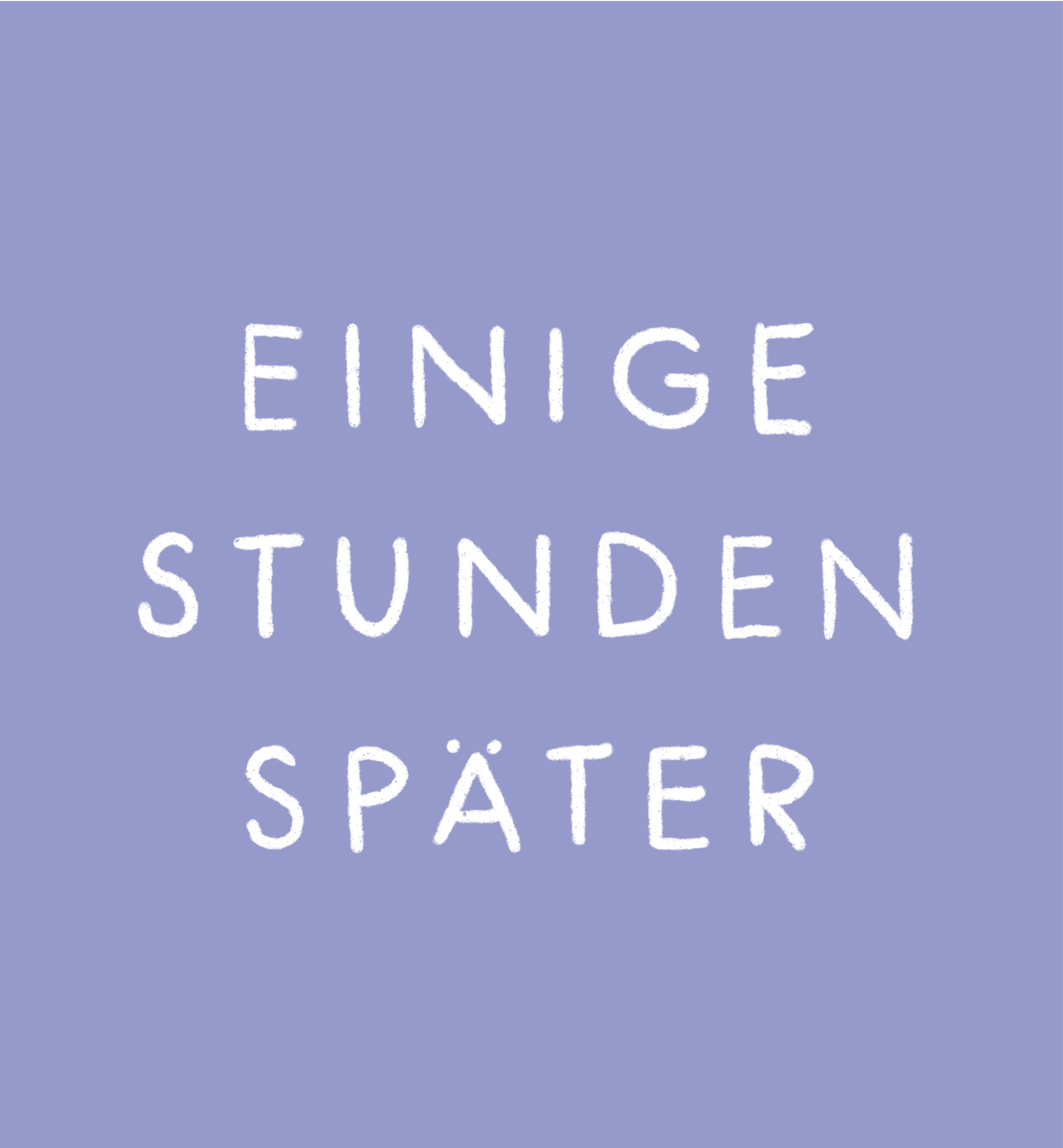}
        \caption{
            The mobile network operator was able to partially restore the network, so communication within the city is possible again, but not to the outside.
        }
        \label{fig:comic24}
    \end{subfigure} 
    \raggedright 
    \begin{subfigure}[t]{\comicwidth} 
        \centering
        \includegraphics[width=\linewidth]{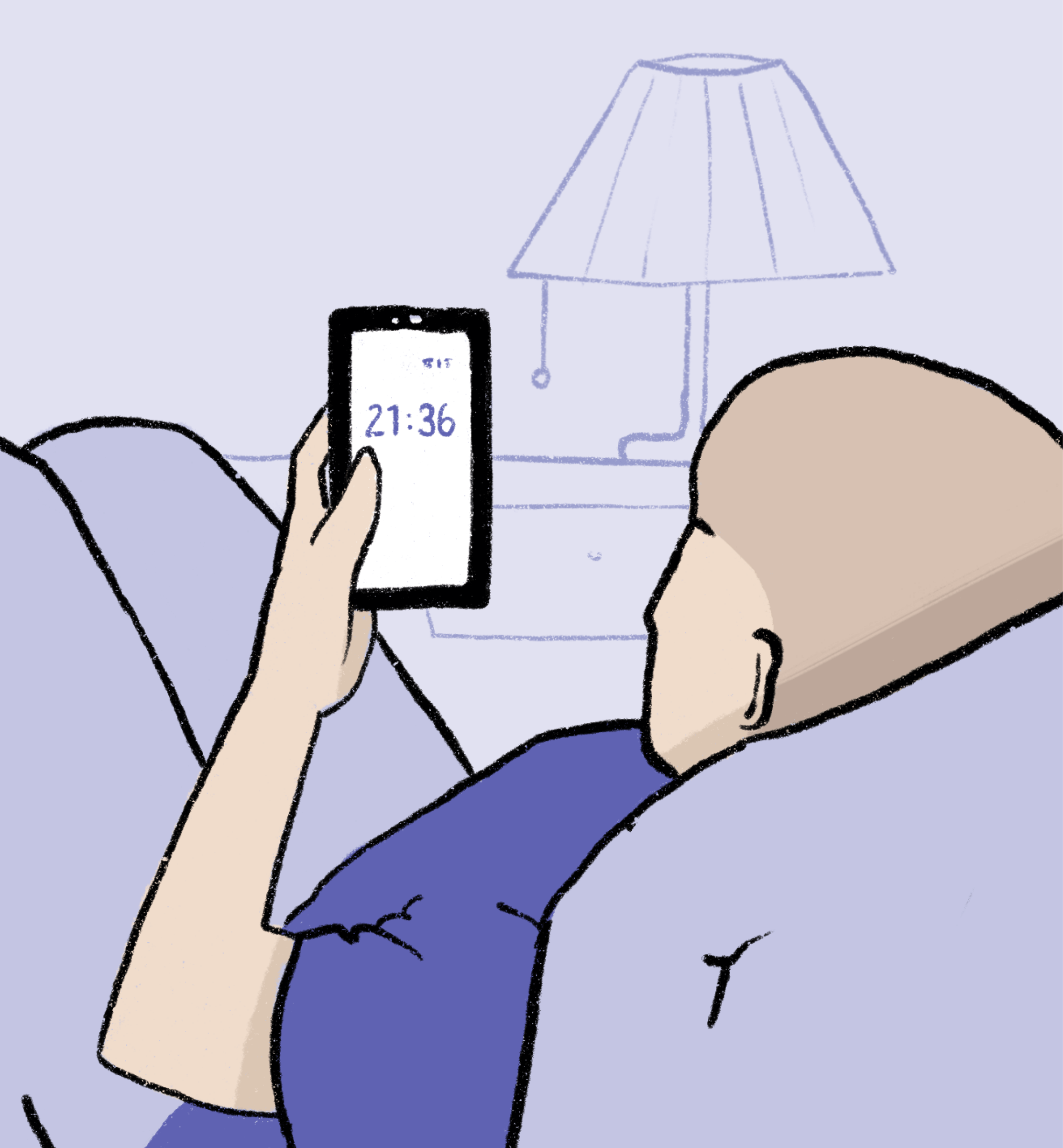}
        \caption{
            At home, you want to use the restored connectivity to inform your friends within the city that you are okay.
        }
        \label{fig:comic25}
    \end{subfigure} 
    \caption{
        \captionheadline{Local Connectivity Model}
        This comic board introduced our participants to the conceptual model of local connectivity in isolated cities.
        The message in Figure~\ref{fig:comic21} translates to "I'm fine!".
        The text in Figure~\ref{fig:comic24} translates to "A few hours later."
    }
    \label{fig:comic2}
    \Description{
        A comic board with five comics.
        Comic 2A shows a hand holding a smartphone with a messenger app opened.
        There is a keyboard visible in the lower half of the screen, and the upper half shows a message that was sent, "I'm fine," but there is an exclamation mark next to the message in its lower-right corner.
        Comic 2B shows the person from the comic board 1 standing on the right with their smartphone in their hand, showing a no-signal symbol in the upper-right corner of the screen.
        The person is looking at a sign, "Free Wi-Fi," that is positioned in the background.
        Comic 2C shows the same person looking at a loudspeaker mounted at the gable of a house, making another announcement.
        Comic 2D shows nothing but text on a blank background, "Several hours later."
        Comic 2E shows the same person as before lying on a couch or bed with their phone in their hand.
        The phone shows the current time, "21:36,".
        There is a lamp in the background.        
    }
\end{figure}}

\subsection{Questionnaire Design}
\label{ssec: method-study}

Section~\ref{sec: survey} holds a full translated transcript of the questionnaire.
It comprises five parts that incrementally familiarize participants with our study scenario.

\paragraph{Part 1: Consent and Screening}
We informed participants about the procedure and goals of the study before they gave their informed consent to participate in our study.
We collected age, gender, education, income, and federal state (Q1-Q5) to steer recruitment towards representativity.
To recruit citizens with prior smartphone experience, we screened whether the participants had lived in a major German city for at least one year and owned a smartphone during that time (Q6).
For eligible participants, we collected the city name and the time the participants lived there (Q7-Q8).
Participants confirmed they had owned and used a smartphone during that time (Q9).
We implemented the city name into the subsequent questions to tailor the questionnaire to the participant, e.g., for participants who had lived in Berlin, the caption of the comic depicted in Figure~\ref{fig:comic11} said: "You are taking a walk in \colorbox{aa}{\textit{Berlin}}."

\paragraph{Part 2: Everyday Smartphone Usage}
We asked the participants to estimate their daily smartphone screen time in hours (Q10), guiding them to find the corresponding Android and Apple iOS statistics.
We studied everyday smartphone usage with positional voting \cite{saari_mathematical_2000}, where participants chose a subset of 23 categories and positioned the chosen app categories by their daily usage frequency (Q11).
There is no generally accepted smartphone app categorization in the literature, so we based our categorization on the Apple App Store \cite{apple_inc_categories_2024}.
We refined it to fit the survey's scope by splitting the \textit{Social Networking} category into \textit{Social Media} and \textit{Messengers,} combining \textit{Photo and Video} and \textit{Graphics and Design} into \textit{Creativity,} introducing categories for \textit{Crisis} and \textit{Web Browsers,} and omitting \textit{Business apps,} \textit{Developer Tools,} \textit{Kids,} \textit{Lifestyle,} and \textit{Safari Extensions}.

\paragraph{Part 3: Crisis Smartphone Usage}
We used a comic board (Figure~\ref{fig:comic1}) to introduce the assumed crisis scenario, an earthquake in the participant's hometown.
We asked the participants to hypothetically answer the remaining survey with this crisis scenario in mind.
Comicboarding \cite{moraveji_comicboarding_2007} is a technique to break down complex information when co-designing with children.
We chose comic boards to explain the study scenario to our non-expert study sample, anticipating the complexity of the local connectivity model in isolated cities.
To study \researchquestion{RQ1}, we chose seven crisis-specific use cases (Q12) and asked which apps participants would use to handle them:
Receiving warnings (U1),
getting safety tips (U2),
contacting emergency services (U3),
looking up information (U4),
communicating to others (U5),
navigating in the city (U6), and
distracting themselves (U7).
We chose these use cases because they were well-established in related work:
\cite{haunschild_perceptions_2022,reuter_increasing_2023,reuter_social_2017} study U1~-~U5, \cite{simko_use_2023,wursching_maintaining_2023} study U6, and \cite{simko_use_2023} studies U7.
Each participant was assigned three of the seven use cases at random to reduce their effort, resulting in \numrange{348}{388} answers per use case (median 364; IQR 19.5).

\paragraph{Part 4: Crisis Smartphone Usage with Island Connectivity}
We showed a second comic board (Figure~\ref{fig:comic2}) to introduce island connectivity into the study scenario:
Internet connectivity is recovered but limited to the participant's city, i.e., communication is only possible within the city, and content is available if hosted on local servers within the city but not from remote servers hosted outside the city.
We showed a video (\SI{38}{\second}) to clarify the scope and limits of island connectivity.
We provided an alternative video description for participants unable to watch videos or encountering technical problems.
We asked participants to answer the remaining questionnaire hypothetically, assuming the crisis was ongoing for several days with local connectivity.
For \researchquestion{RQ2}, we repeated the positional voting category for smartphone usage in isolated cities (Q13).

\paragraph{Part 5: Control Variables and Data Quality}
We collected the following control variables (Q14-Q17):
Prior experience with crises, prior experience with cellular/landline Internet outages, prior activity in relevant emergency forces, e.g., as a firefighter, police officer, or first responder, and the \ac{ATI} scale \cite{franke_personal_2019} with an implemented \ac{IRI} \cite{meade_identifying_2012}.
Finally, participants could indicate technical issues and additional comments in open statements (Q18-Q19).
The survey ended with a \ac{SRSI} UseMe \cite{bruhlmann_quality_2020} (Q20), informing the participants that the survey was an integral part of our research and asking them if we should consider their responses for our studies.

After evaluating the results of our survey, we augmented the findings of \researchquestion{RQ1} and \researchquestion{RQ2} with applicable technology standards \cite{project_3gpp_system_2023} and resilience strategies \cite{abteilung_2_resilience_2022} to construct a smartphone service typology for island connectivity (\researchquestion{RQ3}).

\subsection{Data Analysis}
\label{ssec: method-analysis}

We used the statistical software R 4.4.1 \cite{r_core_team_r_2024} and an alpha level of \num{.05} for the quantitative data analysis.
We corrected for multiple tests with the Bonferroni method \cite{abdi_bonferroni_2007} to evaluate significant associations.
We used the well-known Borda count \cite{mclean_classics_1995} to analyze our participants' positional voting results.
We used inductive coding for the open statements (Q18~-~Q19) and all "Other" responses.
Two researchers independently constructed initial codebooks and merged them through discussions.
Two researchers deductively coded all questions according to that codebook, reaching satisfactory inter-coder reliability (mean Krippendorff's Alpha = \num{.9985}, minimum \num{.997}) \cite{krippendorff_content_2004}.
We solved the remaining coding differences through discussions.
We generally report absolute and relative numbers to classify our results.
We report median values and the \acf{IQR} as these statistics are more robust against outliers.
We report the effect sizes for tests to find significant and relevant associations.
Upon acceptance of this paper, \textbf{we will release a replication package that contains the survey dataset, our evaluation scripts, and the coding system} for readers to verify, reproduce, and replicate our findings.

\subsection{Recruitment and Study Sample}
\label{ssec: method-recruitment}

We chose \gapfish to handle recruitment because they are an \acs{ISO}-certified panel provider cooperating with more than \num{500000} panelists from Germany, Austria, and Switzerland \cite{gapfish_gmbh_gapfish_2024}.
In coordination with \gapfish, we aimed to recruit a representative sample of smartphone users in major German cities based on age, gender, education, and income.
To avoid effort for unfitting participants, \gapfish invited panelists over 18 who owned a smartphone and lived in a German city with more than \num{100000} citizens.
We collected demographic information early in the survey (Q1-Q5) to avoid effort for participants who did not fit the quotas.
Of the \npartials participants who started the survey, \ncomplete participants completed the questionnaire.
After removing participants who
failed \ac{SRSI} UseMe \smalln{64},
failed the \ac{IRI} \smalln{58},
stated invalid education \smalln{4},
stated an invalid major city \smalln{2},
or commented hate speech \smalln{1},
we report $ N = \nfinal $ participants.
\new{Table~\ref{tab:study-sample}} describes our study sample regarding demographic and control variables.
The depicted categories simplify the identification of semantic groups, e.g., younger participants or participants with medium technology affinity.

\paragraph{Representativity}
We recruited a study sample representing smartphone users from major German cities with the following limitations:
Our study sample does not include participants under 18 or over 69 and has a smaller share of participants with low or high education compared to the German population.
GapFish only distinguished male and female panelists in their database, so we could only steer gender representativity regarding men and women.
Still, four participants openly identified as non-binary.
All participants were panelists who had lived in a major German city for at least one year and had owned a smartphone during that time.
We acknowledge these deficits and argue that \textbf{our study sample is representative of adult smartphone users in major German cities} considering the above limitations.

\paragraph{Significant associations}
We found significant associations between our demographic and control variables using Pearson's chi-squared test \cite{pearson_criterion_1992} after correcting for multiple tests with the Bonferroni method \cite{abdi_bonferroni_2007}.
Income and education showed a significant association \associationstats{6}{136.94}{0.001}{0.28} as did age and screen time \associationstats{4}{98.2}{0.001}{0.239}.
These associations represent medium effect sizes, according to Cohen \cite{cohen_statistical_1988}.
We found further significant associations with small effect sizes between
age and education \associationstats{4}{33}{0.001}{0.139},
citizen duration and age \associationstats{4}{67.6}{0.001}{0.119},
citizen duration and screen time \associationstats{4}{27}{0.001}{0.126},
and gender and technology affinity \associationstats{4}{19.7}{0.001}{0.107}.

\subsection{Biases and Limitations}
\label{ssec: method-biases-mitigations}

Naturally, every study has limitations due to the decisions made in the design process.
In the following, we motivate selected design decisions and discuss which biases and limitations they might cause.

\paragraph{Study sample}
We hired an ISO-certified panel provider to recruit and compensate suitable participants.
While recruiting only panelists to complete our questionnaire might introduce a bias against non-panelists, it facilitated steering our study sample towards representativeness for smartphone users from major German cities regarding gender, age, education, and income, thus preventing sampling and selection biases \cite{heckathorn_comment_2011, bethlehem_selection_2010}.
Only participants who had lived in a major German city for at least one year and had used a smartphone during that time could complete our questionnaire.
This screening might introduce a survivorship bias \cite{elston_survivorship_2021} against Germans who have not lived in a major German city for at least one year and German citizens who have not owned a smartphone during that time.
We opted to screen out these groups to align the sample with our study subject, smartphone usage in isolated cities.

\paragraph{Questionnaire design}
We ordered the questionnaire to incrementally familiarize participants with our study subjects to avoid question order bias \cite{schuman_questions_1996}.
Considering the risks of survey time fatigue \cite{lewis_epidemiology_1992}, we determined the dependent variables Q11~-~Q13 as early as possible, showed each participant only three random of the seven use cases in Q11, and included two comic boards \cite{moraveji_comicboarding_2007} and a video before Q13.
On average, it took participants \minsec{12}{38} (median \minsec{10}{26}; \ac{IQR} \minsec{6}{11}) to complete our survey.
We recorded many outliers, with the fastest completion taking \minsec{2}{24} and the slowest completion \qty{10}{\hour}\xspace\minsec{43}{20}.
Following the suggestions of \cite{greszki_exploring_2015}, we did not remove speeding participants.
We implemented a combination of single-choice and multiple-choice selections, number inputs, positional voting, yes/no questions, Likert scales, and open statements to mitigate extreme or neutral response biases and an acquiescence bias \cite{moss_acquiescence_2008}.
Only Q15 used an unbalanced scale, but for this question and, generally, where possible, we provided "No answer" or "Other" options to mitigate a conformity bias \cite{krumpal_determinants_2013}.
We used neutral, non-leading questions and explanations to avoid anchoring biases \cite{tversky_judgement_1974}.
Interviewing the general public about a complex research scenario can introduce a cultural bias \cite{henrich_culture_2015}, so we used comic boards \cite{moraveji_comicboarding_2007} to convey complex information, an explanatory video, and feedback from the pilot study (\cref{ssec: method-ethical}).
We hosted the questionnaire with \textit{LimeSurvey} \cite{limesurvey_limesurvey_2024} to ensure accessibility on various end devices and provided alternative descriptions for the video and comic boards to mitigate technical biases \cite{dillman_internet_2014}.

\paragraph{Data quality}
We opted for an online survey to gather knowledge from a broad study sample and focus on quantitative results.
Ensuring data quality in online surveys can be challenging because participants might be inattentive while filling in the survey, not understand the study subject, or face technical issues.
We anticipated these challenges during in our study design by equipping the questionnaire with an \ac{SRSI} UseMe and an \ac{IRI}, which led to the removal of 122 inattentive participants.
In addition, we implemented comic boarding and an accessible explanatory video with a transcript and description.
While one participant (P162) acknowledged that the study subject "was visionary and hard to empathize with," other participants described the questionnaire as "easy to follow" \smalln{5}, positively mentioning the structure, video, and comics \smalln{5}.
At the end of the survey, we asked participants if they had faced technical issues.
\SI{98.95}{\percent} did not describe technical problems, and five participants (\SI{0.58}{\percent}) had issues sorting the app categories in Q11 and Q13.
Four participants (\SI{0.47}{\percent}) had technical issues with the video, one of which (P402) mentioned that it was "great there was a description below the video."
\new{One participant (P60) proposed expanding Q16 with options for prior and retired emergency forces and hospital staff, which we suggest for future surveys.}

\subsection{Ethical Considerations}
\label{ssec: method-ethical}

We tested our survey in a pilot study with $ N=29 $ participants.
Their valuable feedback led to several improvements, including a shorter survey length, more precise questions, and clarifications.

We adhered to local ethical and legal standards while designing, conducting, and evaluating this study.
Our university's institutional review board reviewed and approved this study.
\gapfish compensated the panelists for their participation in our survey.
As stated in the survey, we did not report participants who failed the \ac{IRI} or self-reported with the \ac{SRSI} UseMe to GapFish.
Despite \gapfish recruiting only adult panelists, two participants reported an age under 18 and were excluded before completing the questionnaire.
We collected demographic information in clusters to protect participants' privacy.
We informed all participants about the survey's purpose and data collection while adhering to the General Data Protection Regulation.
We warned participants that the survey's subject included crises, which might be a sensible topic for them.
Participants agreed to consent before starting the survey, and they could abort their participation at any time without explanations.

\section{Results}
\label{sec: results}

This section reports our survey results, addressing each research question individually.
We supplement our findings with quotes from the open statements to highlight interesting aspects.

\subsection{App Demand for Crisis-Specific Use Cases}
\label{ssec: results-rq1}

\begin{table*}
  \caption{
    \captionheadline{App Demand for Crisis-Specific Use Cases (RQ1)}
    This heat map depicts the seven use cases (U1~-~U7) and the ten most selected app categories.
    For each use case, it shows the percentages for the top three app categories, where \textbf{bold text} highlights the most preferred app category.
    The bottom row $M_c$ shows the mean frequency of how often a category $c$ was mentioned across all use cases.
    The rightmost column $M_u$ shows the mean number of categories selected for use case $u$.
  }
  \Description{
    A heat map with columns for the top 10 apps and rows for the use cases.
    Three sides of the heatmap show additional statistics, i.e., the number of participants that voted on a use case (on the left), the mean number of categories selected per use case (on the right), and the average vote frequency per app category (on the bottom).
  }
  \label{tab:scenario-heatmap}
  \begin{tabular}{lS[table-format=3.2]C{20pt}C{20pt}C{20pt}C{20pt}C{20pt}C{20pt}C{20pt}C{20pt}C{20pt}C{20pt}S[table-format=1.2]r}


    Use case $\setminus$ Category\mytablemark{1} & \multicolumn{1}{l}{$N_u$} & 
    \rotatebox{90}{Crisis}       & \rotatebox{90}{Messengers} & \rotatebox{90}{News}      & \rotatebox{90}{Social Media}  & \rotatebox{90}{Web Browsers}      &
    \rotatebox{90}{Navigation}   & \rotatebox{90}{Weather}    & \rotatebox{90}{Reference} & \rotatebox{90}{Music}         & \rotatebox{90}{Entertainment}     &
    \multicolumn{1}{l}{$M_u$} & None \\

    \midrule 
    
    Warnings (\usecase{U1}) & 380 & 
    \hmctb{65} & \hmct{27} & \hmct{42} & \hmc{14} & \hmc{10} & \hmc{7} & \hmc{16} & \hmc{5} & \hmc{2} & \hmc{2} &
    2.12 & \SI{5}{\percent} \\
    
    Safety tips (\usecase{U2}) & 356 & 
    \hmctb{61} & \hmc{23} & \hmct{39} & \hmc{14} & \hmct{27} & \hmc{7} & \hmc{13} & \hmc{17} &  & \hmc{3} &
    2.28 & \SI{5}{\percent} \\
    
    Emergency call (\usecase{U3}) & 375 & 
    \hmctb{48} & \hmct{37} & \hmc{12} & \hmc{13} & \hmct{13} & \hmc{5} & \hmc{5} & \hmc{4} &  &  &
    1.56 & \SI{16}{\percent} \\
    
    Information (\usecase{U4}) & 348 & 
    \hmct{53} & \hmct{32} & \hmctb{60} & \hmc{25} & \hmc{28} & \hmc{6} & \hmc{12} & \hmc{10} & \hmc{1} & \hmc{1} &
    2.42 & \SI{2}{\percent} \\
    
    Communication (\usecase{U5}) & 364 & 
    \hmct{22} & \hmctb{72} & \hmc{14} & \hmct{35} & \hmc{7} & \hmc{5} & \hmc{5} & \hmc{3} & \hmc{1} & \hmc{1} &
    1.88 & \SI{6}{\percent} \\
    
    Navigation (\usecase{U6}) & 388 & 
    \hmct{20} & \hmct{14} & \hmc{14} & \hmc{6} & \hmc{9} & \hmctb{66} & \hmc{4} & \hmc{5} &  &  &
    1.64 & \SI{3}{\percent} \\
    
    Distraction (\usecase{U7}) & 360 & 
    \hmc{19} & \hmctb{37} & \hmc{21} & \hmct{27} & \hmc{10} & \hmc{5} & \hmc{5} & \hmc{3} & \hmct{27} & \hmc{15} &
    2.42 & \SI{}{9\percent} \\
    \bottomrule 
        $M_c$ & 367.29 & 
    \SI{41}{\percent} & \SI{35}{\percent} & \SI{29}{\percent} & \SI{19}{\percent} & \SI{15}{\percent} &
    \SI{14}{\percent} & \SI{9}{\percent} & \SI{7}{\percent} & \SI{4}{\percent} & \SI{3}{\percent} &
    \\
    \mytablenote{1}{14}{.9\textwidth}{The following app categories did not make the top 10 regarding $M_c$ and were omitted in the heat map for better readability (in descending order according to $M_c$): Utilities, Games, Education, Books, Health \& Fitness, Medical, Travel, Finance, Food \& Drink, Creativity, Productivity, Shopping, Sports.}\\
  \end{tabular}
\end{table*}

Our first goal is to understand which apps citizens intuitively prefer to use for crisis-specific use cases (\researchquestion{RQ1}).
Notably, the list of use cases covered by our study is not exhaustive, and other use cases might yield different results.
We decided to include U1~-~U7 because related work \cite{haunschild_perceptions_2022,reuter_increasing_2023,reuter_social_2017,simko_use_2023,wursching_maintaining_2023} studied similar use cases but not with a focus on citizens of major German cities.

Immediately after introducing the crisis scenario with the comic board depicted in Figure~\ref{fig:comic1}, we showed each participant three of the seven use cases.
We asked them which apps they would use in these scenarios.
Table~\ref{tab:scenario-heatmap} visualizes the results with a row per use case (U1~-~U7).
On average, participants selected \num{2.03} app categories per use case (median \num{1.67}; IQR \num{1.67}).
The most often selected app category across the $ \nfinal \cdot 3 = 2571 $ votes was \appcategory{crisis apps,} selected \num{1066} times and scoring highest for receiving warnings (\usecase{U1}) and getting safety tips (\usecase{U2}).
For the remaining use cases, participants preferred other apps over crisis apps:
\appcategory{News apps} for looking up information (\usecase{U4}),
\appcategory{messengers} for communication (\usecase{U5}) and distraction (\usecase{U7}),
and \appcategory{navigation apps} for navigating in the city (\usecase{U6}).
While crisis apps scored highest for contacting emergency services (\usecase{U3}), many participants preferred emergency calls for this use case, selecting "No app" or using the "Other" option to describe telephony (\SI{16}{\percent}).
\new{Overall, this highlights that it takes more than crisis apps to serve all crisis-specific use cases and that users prefer using general-purpose apps in crisis scenarios.}

\subsection{App Demand on Islands}
\label{ssec: results-rq2}

\begin{table}
    \centering
    \caption{\captionheadline{App Demand on Islands (RQ2)}
    This table compares how apps ranked in normal operation (Table~\ref{tab:rq2-normal}) and island operation (Table~\ref{tab:rq2-island}).
    Dotted arrows indicate trends of two ranks or more, and bold blue arrows highlight \textcolor{myblue}{\textbf{strong trends}}.}
    \Description{
    There are two tables each showing positional voting results:
    On the left, there are the results of Q11 as a list of apps, and on the right, there are the results of Q13 as a list of apps.
    There are arrows from the left to the right table.
    Flat arrows are in black color with a dotted line and there are steep arrows in a bold blue line:
    Crisis apps from Rank 21 to Rank 3, Shopping apps from Rank 6 to Rank 19, and Travel apps from Rank 16 to Rank 21.}
    \includegraphics[scale=1]{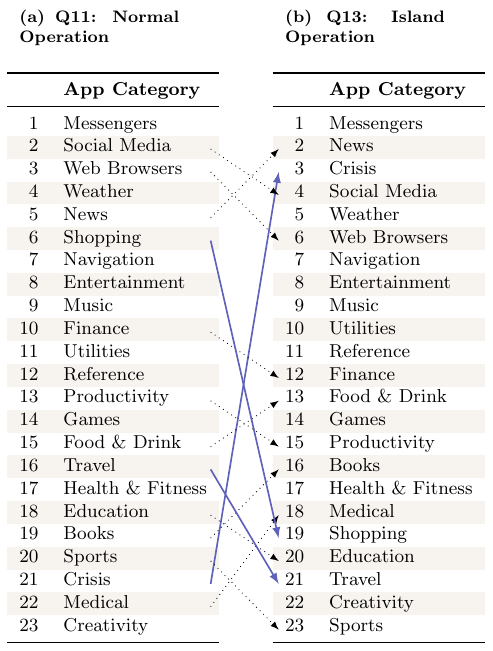}
    \label{tab:rq2}
    \label{tab:rq2-normal}   
    \label{tab:rq2-island}  
\end{table}

\begin{figure*}
    \centering
    \begin{subfigure}[t]{.24\textwidth}
        \centering
        \includegraphics[scale=1]{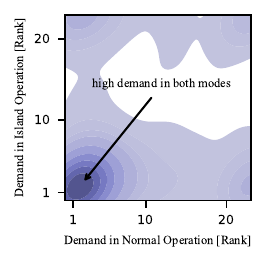}
        \caption{
          Messengers.
        }
        \label{fig:trend-messenger}
    \end{subfigure}
    \hfill
    \begin{subfigure}[t]{.24\textwidth}
        \centering
        \includegraphics[scale=1]{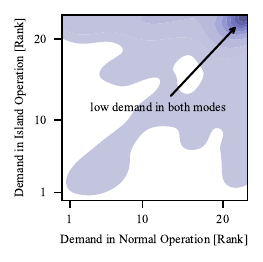}
        \caption{
          Creativity apps.
        }
        \label{fig:trend-creativity}
    \end{subfigure}
    \hfill
    \begin{subfigure}[t]{.24\textwidth}
        \centering
        \includegraphics[scale=1]{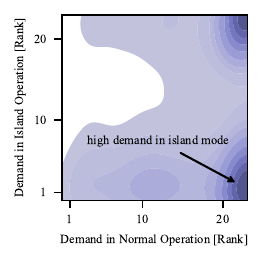}
        \caption{
          Crisis apps.
        }
        \label{fig:trend-crisis}
    \end{subfigure}
    \hfill
    \begin{subfigure}[t]{.24\textwidth}
        \centering
        \includegraphics[scale=1]{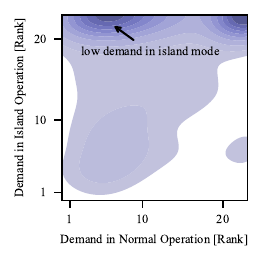}
        \caption{
          Shopping apps.
        }
        \label{fig:trend-shopping}
    \end{subfigure}
    \caption{
        \captionheadline{Positional Voting Distribution for Apps with Strong Trends}
        Each figure compares the vote distribution of the app demand in normal operation (Q11) and island operation (Q13) in a bivariate \acf{KDE} plot \cite{terrell_variable_1992}.
        The annotated density peaks underscore strong trends identified in Table~\ref{tab:rq2}.
        For instance, the soaring demand for crisis apps in island operation compared to normal operation (Figure~\ref{fig:trend-crisis}) causes a density peak in the lower-right corner, as participants voted for crisis apps near Rank 23 (low demand) in Q11 and near Rank 1 (high demand) in Q13.
        Participants who prefer not to use crisis apps in normal nor island operation account for the smaller density peak in the upper-right corner.
    }
    \label{fig:rq2-trends}
    \Description{
        There are four side-by-side kernel density estimation plots.
        Plot A has one density peak at the lower-left corner annotated with an arrow and text "high demand in both modes."
        Plot B has one density peak at the upper-right corner annotated with an arrow and text "low demand in both mdoes."
        Plot C has two density peaks: One weaker peak at the upper-right and a stronger peak at the lower-right corner that is annotated with an arrow and text "high demand in island mode."
        Plot D has two density peaks: One weaker peak at the upper-right corner and a stronger peak at the upper-left corner that is annotated with an arrow and text "low demand in normal mode."
    }
\end{figure*}

For \researchquestion{RQ2}, we focus on two app types:
Apps that our study sample ranked high in island operation and apps with strong trends from island operation to normal operation.
Our questionnaire featured two positional voting questions to learn about the participants' app demands in normal operation (Q11) and island operation (Q13).
In this context, \textit{normal operation} refers to a non-crisis scenario with global Internet connectivity, and \textit{island operation} refers to the scenario where only island connectivity within the island is feasible.
Notably, many methods exist to compute ranks from positional votes.
We used the Borda count \cite{mclean_classics_1995}, a widely recognized and straightforward method, as it generally produces a broadly acceptable consensus based on individual rankings.
Since using a different method could yield different results, we also publish our raw data to enable other researchers to apply alternative methods.
Table~\ref{tab:rq2} compares the positional voting results of all apps in normal and island operation, and Figure~\ref{fig:rq2-trends} takes a closer look at the vote distribution for apps with strong trends.

\paragraph{Apps with Unchanged Demand In Island Operation}
Ten apps scored in the same rank for normal operation (Table~\ref{tab:rq2-normal}) and island operation (Table~\ref{tab:rq2-island}) or with a difference of one rank.
Table~\ref{tab:rq2} does not show arrows for these apps for better readability.
\appcategory{Messengers} scored highest in both operation modes.
Most participants ranked messengers near Rank 1 in Q11 and Q13, resulting in a density peak at the lower left corner of Figure~\ref{fig:trend-messenger}.
From RQ1, we understand that citizens prefer messengers to communicate with other people for communication and distraction.
However, island connectivity limits communication to people in the city.
For example, P457 commented, "As I do not have any friends or acquaintances in Düsseldorf, [\dots] I would not want to communicate with anyone within the city in that scenario."
We found similar unchanged but low demands for \appcategory{creativity} apps, causing the density peak in the upper right corner of Figure~\ref{fig:trend-creativity}.
Most participants who voted for creativity apps in Rank 23 in Q11 voted for them in Rank 22 in Q13.
Apps related to weather, navigation, entertainment, music, utilities, reference, games, and health \& Fitness showed a similarly unchanged demand.

\paragraph{Apps with Increased Demand in Island Operation}
Five apps scored higher for island operation than normal operation.
As a crisis-specific medium, \appcategory{crisis apps} climbed 18 ranks from normal operation to island operation.
This is reflected in Figure~\ref{fig:trend-crisis}, where a density peak in the bottom-right corner indicates that crisis apps were voted near Rank 1 in island operation but near Rank 23 in normal operation.
However, the second peak in the top-right corner indicates a share of our study sample with low demand for crisis apps regardless of the operation mode.
We observe a trend of participants moving from Ranks 6~-~10 in normal operation to Ranks 1~-~5 in island operation, visible as a white area in Figure~\ref{fig:trend-crisis}.
This trend suggests that participants who occasionally use crisis apps in their daily lives use them more frequently on islands, which aligns with our findings regarding \researchquestion{RQ1} reporting high demands for crisis apps to receive warnings and safety tips, contact emergency services, and seek information.
We found weak positive demand changes for apps related to news, food \& drink, books, and medical.

\paragraph{Apps with Decreased Demand in Island Operation}
Eight apps scored lower for island operation than normal operation.
The demand for \appcategory{shopping apps} plummeted the most, described by the density peak in the upper-left corner of Figure~\ref{fig:trend-shopping}.
Most participants who voted for shopping apps in the top-10 ranks in Q11 moved to the lowest ranks in Q13.
Similarly, shopping apps show a secondary peak as some participants showed no demand for shopping apps at all.
Social media, web browsers, finance, productivity, education, travel, and sports apps showed a similar demand decrease.

\paragraph{Apps with High Demand in Island Operation}
Three apps complement the top-5 ranks for island operation besides the aforementioned messengers and crisis apps.
\appcategory{News apps} climbed from Rank 5 in Q11 to Rank 2 in Q13.
From \researchquestion{RQ1}, we understand that our participants prefer news apps for looking up crisis-relevant information as well as receiving warnings and safety tips.
The increased demand in island operation underscores the importance of news apps for islands.
\appcategory{Social media} descends from Rank 2 in normal operation to Rank 4 in island operation, as the use of social media is limited to the island in island connectivity.
The results of \researchquestion{RQ1} show that participants primarily prefer social media for communication and distraction.
We find different demands for messengers and social media among our study sample, which confirms our design decision to split Apple's \textit{Social Networking} app category into \textit{Social Media} and \textit{Messengers,} as the demands of these app categories seem to differ on islands.
\appcategory{Weather apps} can play a crucial role in crisis-struck isolated areas, especially considering natural disasters.

\subsection{Smartphone Services Typology for Island Connectivity}
\label{ssec: results-rq3}

\begin{figure}
    \centering
    \includegraphics[width=\columnwidth]{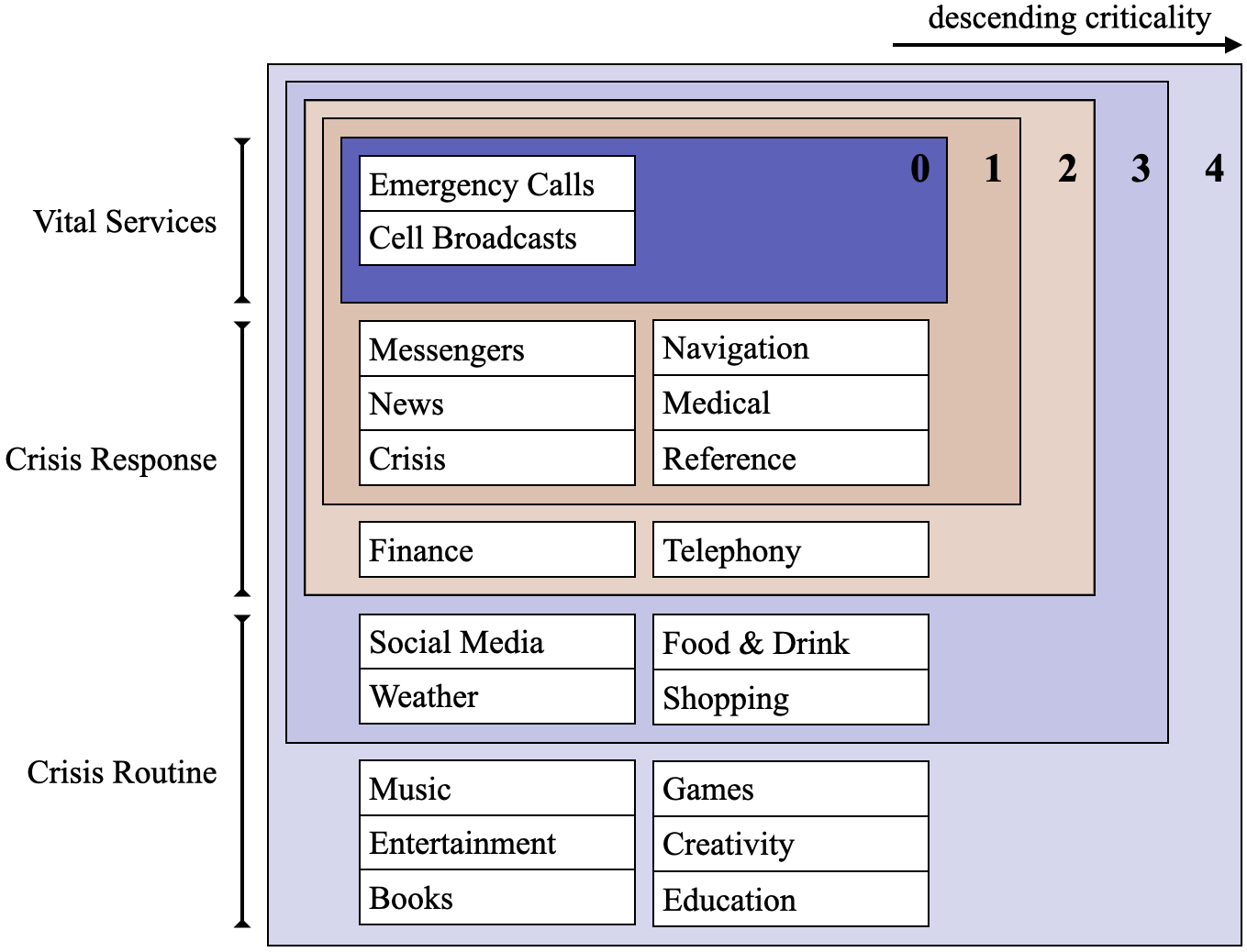}
    \caption{
        \captionheadline{Smartphone Service Typology for Island Connectivity (RQ3)}
        This prioritization organizes smartphone services in tiers based on their criticality for islands.
        Tier 0 contains vital services for islands, Tiers 1~-~2 focus on immediate crisis response, and Tiers 3~-~4 focus on crisis routine.
    }
    \label{fig:rq3}
    \Description{
        There are five planes on top of each other.
        The basis is a white plane labeled "5" with music, entertainment, books, games, creativity, and education in it.
        Next is a light-beige plane labeled "4" with social media, weather, food and drinks, and shopping in it.
        An arrow on the left indicates that planes 3 and 4 belong to the "crisis routine."
        On top is a beige plane labeled "2" with finance and telephony in it.
        On top is a light-blue plane labeled "1" with messengers, news, crisis apps, navigation, medical, and references in it.
        An arrow on the left indicates that planes 1 and 2 belong to the "crisis response."
        The topmost plane is colored blue and labeled "0," with emergency calls and cell broadcasts in it.
        An arrow on the left indicates that plane 0 belongs to "vital services."
    } 
\end{figure}

\begin{figure*}
    \centering
    \begin{subfigure}[t]{.47\textwidth}
        \centering
        \includegraphics[scale=.72]{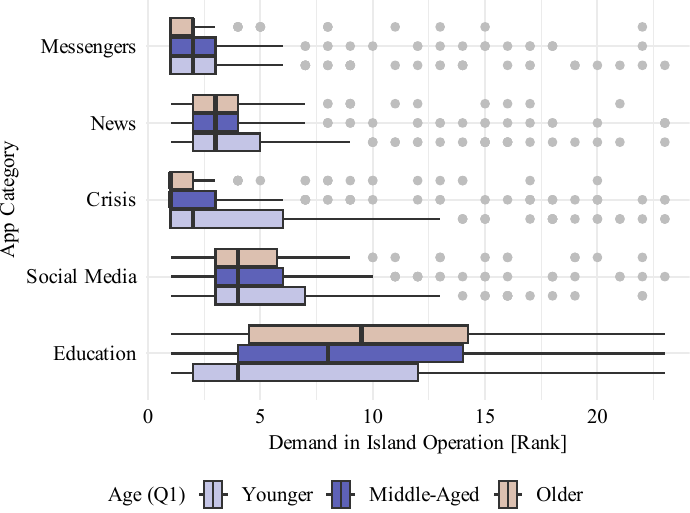}
        \caption{
          \new{Rank distribution of groups based on age (Q1).}
        }
        \label{fig:rankdist-age}
    \end{subfigure}
    \hfill
    \begin{subfigure}[t]{.47\textwidth}
        \centering
        \includegraphics[scale=.72]{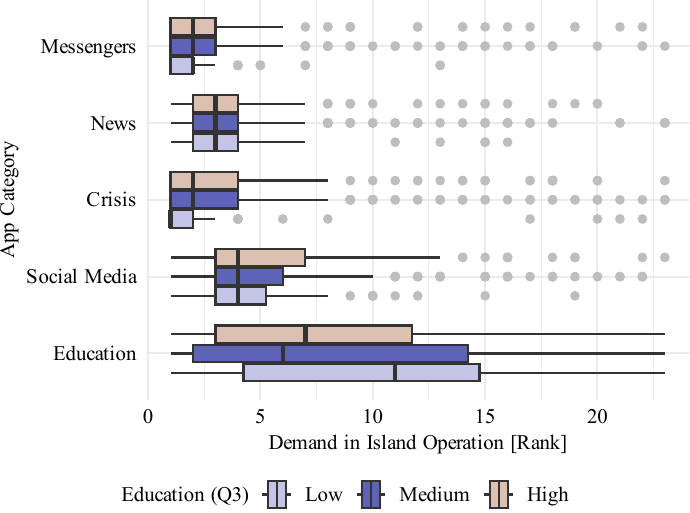}
        \caption{
          \new{Rank distribution of groups based on education (Q3).}
        }
        \label{fig:rankdist-education}
    \end{subfigure}
    \caption{
        \new{
        \captionheadline{Rank Distribution Tendencies of Semantic Groups}
        This figure compares the app demand of semantic groups based on age (Figure~\ref{fig:rankdist-age}) and education (Figure~\ref{fig:rankdist-education}).
        Each figure features box plots to show the app demand in island operation (Q13) of a group (e.g., younger adults or participants with high education) for a selected app (e.g., messengers or crisis apps).
        Table~\ref{tab:rankdist-all} gives an overview of all considered variables.}
    }
    \label{fig:rankdist-selected}
    \Description{
        There are two side-by-side plots with box plots for five categories (Messengers, News, Crisis, Social Media, Education). Plot A has one box plot per app category and age group (Younger, Middle-Aged, Older). Plot B has one box plot per app category and education group (Low, Medium, High). In both plots, the median ranks for messengers, news, crisis apps, and social media show only minimal variance. For education apps, there is more variance: The median rank of education apps among younger adults is Rank 4, while middle-aged and older adults ranked them with medians of Rank 8 and Rank 9. Similarly, in Plot B, the median rank of education apps among low-educated participants was Rank 11, while medium-educated and high-educated participants ranked in Rank 6 and Rank 7.
    }
\end{figure*}

Our contribution to \researchquestion{RQ3} is a smartphone service typology for island connectivity.
In this context, \textit{smartphone services} generalize the functionality provided by smartphone apps and services of the cellular network.
We construct this typology by combining the fieldwork-informed knowledge from our survey with practice-informed design knowledge from authorities \cite{abteilung_2_resilience_2022,projekt_415_los_1_zwiback-studie_2023} and applicable technology standards \cite{project_3gpp_system_2023}.
Figure~\ref{fig:rq3} visualizes how the typology classifies smartphone services into five tiers, prioritizing how critical the included services' availability is on islands.
\new{The typology is based on our findings for research questions \researchquestion{RQ1} and \researchquestion{RQ2}, which we augment with demographic tendencies observable in our study sample. Figure~\ref{fig:rankdist-selected} depicts a selection of tendencies and Table~\ref{tab:rankdist-all} shows the median ranks of all categories by semantic groups.}
While Tiers 0 and 1 focus on the immediate crisis response, distinguishing paramount smartphone services and crucial apps,
Tiers 2 and 3 consider cities in crisis routine, identifying required enablers and apps for day-to-day crisis life.
Tier 4 is about citizens' well-being during crises, discussing the potential of nice-to-have apps during crisis recovery.
In the following, we discuss each tier individually:

\paragraph{Tier 0: Crisis-Critical Cellular Services}
Making \appcategory{emergency calls} and receiving \appcategory{cell broadcasts} are central pillars of the German federal network agency's resilience strategy \cite{abteilung_2_resilience_2022}.
European and federal laws regulate the availability of emergency calls and cell broadcasts, e.g., users can make emergency calls without money on their account and use another \ac{MNO}'s network \cite{moller_emergency_2015}, highlighting its crucial importance for crisis response.
The results of our survey support this assessment, as despite not having asked about it explicitly, \SI{16}{\percent} of the participants reported they would not use smartphone apps but telephony to make an emergency call.
The services of Tier 0 should be the absolute minimum that every \ac{MNO} maintains on islands.

\paragraph{Tier 1: Services for Immediate Crisis Response}
We found that \appcategory{messengers} are the most demanded app category on islands, covering crisis-specific communication and distraction use cases.
Communicating with close ones affected by the crisis can alleviate stress and avoid extensive movement through the crisis-struck area \cite{stute_reverse_2017,alizadeh_impacts_2023}.
\new{Messengers show similar rank distributions for all considered variables (Table~\ref{tab:rankdist-all}).
All semantic groups in our study sample showed similarly high demand for messengers \medianrank{2}, e.g., see Figure~\ref{fig:rankdist-selected} to compare rank distributions of groups based on age and education.}
Our participants preferred \appcategory{news apps} to look up information.
While providing access to global or generally external newspapers is hard, if not infeasible, at least local newspapers should be available on islands.
As the name suggests, \appcategory{crisis apps} are dedicated to crisis scenarios, including islands.
While we classify cell broadcast as a Tier 0 service, which is paramount to broadcasting urgent information to all citizens, crisis apps can provide more detailed warnings and additional information about the crisis \cite{hauri_comparative_2022}.
\new{Similarly to messengers, the rank distribution of news and crisis apps shows minimal variance between semantic groups with medians around Rank 3.}
\appcategory{Navigation apps} can help citizens navigate isolated areas depending on the crisis.
\new{Local map data can be hosted on the island, and local crisis scenarios usually do not affect GPS satellites. Furthermore, should GPS be unavailable, e.g., due to a targeted cyber attack, the radio-based positioning of 6G can be utilized \cite{behrevan_positioning_2024}.}
These apps become especially useful when augmented with crisis-relevant details, e.g., available and blocked roads, water sources, or places with Wi-Fi connectivity \cite{haesler_connected_2021}.
\appcategory{Medical apps} can be vital for patients relying on dedicated apps, e.g., monitoring insulin for diabetes patients.
Maintaining the availability of medical apps can prevent crisis-unrelated emergencies of patients, avoiding additional stress on first responders \cite{hussain_landscape_2015}.
\new{We observed several tendencies regarding medical apps in our study sample:
Participants with a short or medium residence time \medianrank{8} and younger participants \medianrank{9} showed lower demands for medical apps compared to participants with long residence times \diffrank{-6}, and older participants \diffrank{-4}, respectively.}
\appcategory{Reference apps} can provide crisis-relevant information for citizens, e.g., how to filter water or prepare food without electricity \cite{lakeman_practical_2017}.

\paragraph{Tier 2: Services with Critical Purpose}
\appcategory{Finance apps} give citizens access to their finances, which is a critical infrastructure in many countries \cite{weber_mapping_2023}.
Our study does not find an increased demand for finance apps on islands.
Still, payment apps and, ultimately, the critical infrastructure of finances must be available on islands to enable an organized day-to-day crisis life, which poses substantial challenges as we elaborate in \cref{ssec: discussion-implications}.
\new{We observe higher demand for finance apps in high-income and medium-income participants \medianrank{7, both} than low-income participants \diffrank{-4}.}
\appcategory{Telephony} enables citizens to make phone calls within islands and is another enabler of crisis routines \cite{simko_use_2023}.

\paragraph{Tier 3: Services for Crisis Routine}
Our survey shows a high demand for \appcategory{social media} and \appcategory{weather apps} on islands.
Since the terrorist attacks of 9/11, the \ac{HCI} and \ac{CSCW} communities have observed an increasing use of social media in crises \cite{reuter_social_2018}.
Related studies investigated use cases for social media during crisis response, e.g., to organize community efforts \cite{starbird__2011} or to use media from crisis-struck regions as input for situational awareness \cite{zander_aware_2023}.
As such, individual use cases for social media during crisis response exist even though social media's primary purpose is in crisis routine.
\appcategory{Weather apps} have become an integral part of our everyday lives, not necessarily opening the weather app itself but reading snippets of weather information on lock screens, info panels, or menu bars.
\appcategory{Food \& drink} and \appcategory{shopping apps} might not seem as crucial for crisis routines as other apps.
However, food supply is a critical infrastructure in Germany, and shopping apps can mean shops offering crisis-relevant equipment.

\paragraph{Tier 4: Services for Crisis Well-Being}
\appcategory{Music apps} play a central role for citizens seeking distraction from the crisis \new{\cite{ziv_2022_music}}.
\new{In our sample, younger adults had a higher demand for music apps during crises \medianrank{5} than middle-aged adults \diffrank{-2} and older adults \diffrank{-7}.}
\appcategory{Entertainment apps} for TV shows or movies, \appcategory{books} for reading, and \appcategory{games} for gaming are further sources of distraction \cite{hardayati_implementation_2019}.
While our survey found little demand for \appcategory{creativity apps} and \appcategory{education apps} on islands, their availability can benefit crisis routine.
\new{Among our study sample, we observe high demand for education apps among younger participants and participants with low technology affinity \medianrank{4, both} compared to older participants \diffrank{-5.5} and participants with high technology affinity \diffrank{-4}, respectively.
Further, we observed a tendency linking lower demand for education apps with lower education (Table~\ref{tab:rankdist-all}).}

\paragraph{Limitations of this Classification}
The above typology omits some apps for different reasons.
\appcategory{Web browsers} stand out from all other apps from a technical perspective, considering that web browsers give users access to the World Wide Web with a virtually infinite number of servers \cite{mccauley_extracting_2023}.
Notably, our system model does not distinguish app-based and web-based Internet traffic.
Island operation assumes that app servers on the network edge are available for users even when an island is disconnected from the outside Internet. The same holds for web servers, i.e., web pages with a web server hosted on the island will be available while web pages hosted on remote web servers will not.
Website operators can change their backend communication infrastructure similar to app operators, enabling the use of their websites on islands.
Therefore, the availability of apps and websites is somewhat equivalent on islands, making web browsers orthogonal to our classification.
However, considering the enormous number of websites and the limited capacity of edge servers, island-ready websites will be feasible only for a few website operators and most interesting to local content providers such as local newspapers or communities.
\appcategory{Productivity apps} and \appcategory{utilities} usually work offline and would be available to island users if downloaded before the crisis.
The applicability of \appcategory{travel} and \appcategory{sports apps} with island connectivity is limited as no information from the outside Internet is available.

\subsection{Qualitative Findings}
\label{ssec: results-qualitative}

\paragraph{App Prioritization}
Three participants commented that not all app categories are equally important in crisis scenarios.
One participant (P136) commented, "Several considered apps will never have a practical use nor will they be used during a crisis."
Similarly, P902 found the available app categories "too many and in parts absurd," asking, "Why should I read a book or listen to music during an earthquake?".
This statement contrasts our observed need for distraction during crises, as \SI{27}{\percent} preferred \appcategory{music apps} for distraction (\researchquestion{RQ1}).
In addition, \appcategory{music} stayed on Rank 14 while \appcategory{books} gained three ranks from normal to island operation (\researchquestion{RQ2}).
P475 demanded that "priorities should be set, since, e.g., games and apps for ordering food should be considered low-priority/useless," which indicates demand for the smartphone service typology we propose for island connectivity (\researchquestion{RQ3}).

\paragraph{Non-App Technologies}
Thirteen participants (\SI{1.52}{\percent}) used the "Other" option of Q12 to indicate they preferred telephony over apps for the use cases of emergency calls \smalln{7}, information \smalln{1}, communication \smalln{4}, and distraction \smalln{1}.
This is complemented by the comments of four participants who "only use [their] smartphone for phone calls, SMSes, and utilities" (P148).
Similarly, two participants criticize the overreliance on apps and smartphones, commenting that "one should never solely depend on smartphones" (P121) and regretting the shift away from traditional landline phones that were equipped to function during power outages (P742).
These comments suggest prioritizing \appcategory{telephony} higher than Tier 2 in the smartphone service typology (\researchquestion{RQ3}), potentially as a vital Tier 0 service.
However, the small share of participants preferring telephony over general-purpose apps and the large share preferring general-purpose apps during crisis response motivate us to prioritize Tier 1 apps over telephony services.

\paragraph{Scenario Realism}
Three participants commented that the earthquake scenario is implausible in Germany, to which one participant (P384) added that "flooding would be a better-suited example."
This indicates that geographic factors should be considered when prioritizing apps for crisis scenarios.
Six participants found the described island operation doubtful; e.g., one participant (P69) stated, "The described scenario is wrong because the Internet does not work locally."
These comments suggest that some participants were inattentive while reading the survey or did not understand the study subject, as the main idea of island connectivity is maintaining local Internet connectivity, which the questionnaire explains elaborately.

\section{Discussion}
\label{sec: discussion}

This section positions our findings relative to prior work and discusses the practical implications for operators, developers, and authorities towards island-ready \ac{6G} communication.

\subsection{Contextualizing Our Findings}
\label{ssec: discussion-findings-context}

Considering our German sample, we found demand for other apps than crisis apps for crisis-specific use cases (\researchquestion{RQ1}).
While crisis apps were most demanded for warnings (U1), safety tips (U2), and emergency calls (U3),
participants preferred messengers for communication (U5) and distraction (U7), news apps for looking up information (U4), and navigation apps for navigation (U6).
These results support the findings of related work that dedicated crisis apps enjoy an increasing relevance for warnings, safety tips, and emergency calls and that crisis apps are perceived as more helpful sources for crisis communication than other channels.
While in 2019, social media was still perceived as a quite or very helpful (\SI{50}{\percent}) information source \cite{haunschild_sticking_2020}, this perception decreased considerably until 2021 (only \SI{28}{\percent}) in line with an increasing fear of false information disseminated through social media (\SI{54}{\percent} in 2019 vs. \SI{76}{\percent} in 2021) and concerns about their reliability during emergencies (\SI{38}{\percent} in 2019 vs. \SI{48}{\percent} in 2021) \cite{reuter_increasing_2023}.
Furthermore, in both cited studies, messengers were the most used type of social media to search for and share information.
This change in social media perception and use aligns with our findings that not social media but crisis, messengers and news apps were mentioned among the most important sources for warnings, safety tips and information.
However, this work's novel contribution compares use cases with general-purpose apps \cite{tan_mobile_2017}, highlighting the critical complementary roles of news and messenger apps.

In isolated areas with island connectivity (\researchquestion{RQ2}), German citizens report high demands for messengers, news apps, crisis apps, and social media apps.
The demand for crisis apps was exceptionally high compared to normal operation, while the demand for shopping and travel apps dropped the most.
Thus, being informed about the ongoing event alongside social exchange via messengers and social media constitutes essential smartphone-based activities in an island setting, emphasizing the need to provide them with decentralized local cellular connectivity \cite{sterz_energy-efficient_2023,heise_optimized_2022}.
We assume that leisure activities such as entertainment, music, and gaming are still among important coping and distraction strategies, which, however, the widely available types of gaming, personal computers, and television devices \cite{kaufhold_implementation_2022} might fulfill.
Overall, while prior user research examined the impact of infrastructure outages \cite{fekete_here_2021,thieken_performance_2023}, social media \cite{reuter_impact_2019,wang_understanding_2022}, and warnings apps \cite{fischer-presler_protection-motivation_2022,tan_usability_2020} in crises, this work extends upon them by considering the concept of island connectivity.

We develop a smartphone service typology for island connectivity (\researchquestion{RQ3}) that combines smartphone apps and cellular services into tiers ordered by the criticality of their availability.
Our typology gives recommendations to \acp{MNO}, app operators, and authorities, e.g., to maintain emergency calls and cell broadcasts before ministering telephony and to prioritize the availability of messenger apps on islands over social media apps.
These align with the resilience strategy of the German federal network agency that demands \ac{MNO} to ensure essential services with the highest priority \cite{abteilung_2_resilience_2022}.
They also describe that \acp{MNO} should prioritize smartphone apps when the cellular network faces limited capacity, which aligns with applicable resilience definitions \cite{khaloopour_resilience-by-design_2024,hollick_emergency_2019}.
We complement these general suggestions with a concrete typology that enables stakeholders to prioritize smartphone services.

\subsection{Design Implications}
\label{ssec: discussion-implications}

We combined the crowd-sourced understanding gained from our survey with practice-informed knowledge of applicable standards and federal guidelines to derive a smartphone service typology.
With the ongoing development of \ac{6G}, local connectivity on islands is still a vision, and many challenges remain to be solved by \acp{MNO}, smartphone app developers, and authorities.
While the vision of island connectivity aligns well with the decentralized web \cite{raman_challenges_2019} and local-first software \cite{kleppmann_local-first_2019}, we emphasize that \textbf{the realization of islands and island connectivity poses non-trivial challenges for many stakeholders and requires a substantial review of current standards}.
Proposing solutions to these challenges is outside the scope of this paper, but we outline design implications for selected involved parties in the following.
With these, we support stakeholders in making user-centric design decisions during the realization of island connectivity as a part of 6G communication.

\paragraph{Towards Island-Ready 6G}

The vital requirement to realize the vision of island connectivity in crisis-struck areas is the \designcriterion{distribution} \designcriterion{of the core network.}
Equipping every major city with a core network replica would improve resilience and maximize synchronization efforts and costs \cite{corici_shortcuts_2024}.
A trade-off could be to equip each metropolitan area with a core network replica or to co-locate each Internet gateway with a complete core network instance.
The cellular network should be able to \designcriterion{transition into and out of island operation.}
When an area is disconnected from the outside Internet, the core network replica should detect the isolation and enter island operation.
Alternatively, Internet connections could follow a local-first approach \cite{kleppmann_local-first_2019} and access fallback to the Internet only when the requested content is unavailable on the local network edge.
The \designcriterion{availability of critical smartphone services} should be prioritized depending on the crisis and the cellular network's capacity.
Ideally, users could use all smartphone apps on islands, but \acp{MNO} might have to prioritize available services due to limited capacity.
In that case, the smartphone service typology (\researchquestion{RQ3}) suggests making Tier 0 cellular services the main priority.
At the same time, the availability of apps in Tiers 1~-~4 should be prioritized with descending priority, e.g., enabling citizen-to-citizen communication with messenger apps before enabling entertainment apps whose primary use case is distraction.

\paragraph{Towards Island-Ready Smartphone Apps}
\new{Our findings regarding \researchquestion{RQ1} indicate that users would like to continue using general-purpose apps during crises, in addition to or instead of dedicated crisis apps.
Therefore, apps should be available in island-ready 6G without requiring users to download a separate app version for island operation.}
Ideally, apps should be able to \designcriterion{transition into island} \designcriterion{operation} immediately when the area is disconnected from the outside Internet, which likely requires changes to the communication design of how apps connect to remote servers.
Multi-access edge computing and local-first communication patterns \cite{kleppmann_local-first_2019} could play a central role in this endeavor, e.g., the app on a user's smartphone could connect to an app server on the local network edge, which (a) forwards traffic to a remote app server in normal operation or (b) enables local use of the app in island operation.
\new{For island operation, one specific challenge is ensuring security and privacy, as many apps rely on centralized trust infrastructure and identity providers, making it important to explore \designcriterion{local or decentralized authentication methods.}
For messengers, which were most important to our participants overall, stronger emphasis needs to be placed on ensuring confidential \ac{E2EE} communication without requiring access to centralized authentication systems.
Considering that the challenges of local connectivity will impact user interaction, \designcriterion{apps in island operation should be user-friendly,} i.e., how to announce the transition into island operation, how to establish trust locally, and how to explain to users the limitations of island connectivity for that app \cite{haesler_connected_2021}.}
When the Internet connection is recovered, apps should be able to \designcriterion{transition out of island operation,} which likely requires the application of disruption-tolerant networking principles \cite{li_mops_2009,gao_user_2011,gao_supporting_2011}.

\paragraph{Towards Island-Ready Regulations}

We envision that smartphone apps and 6G will be island-ready, supporting citizens during crisis response.
Enabling island users to download smartphone apps requires a \designcriterion{distributed crisis app store} on each island.
To this end, app stores could be distributed similarly to app servers, providing a selection of apps and content suitable for potential crisis scenarios.
\new{Distributed app stores would further facilitate considering geographic factors when preparing a set of crisis-relevant apps for crisis scenarios.}
The responsibility for these app stores is an open problem, as is their content.
On the one hand, the threat scenarios differ for each city, so the criticality of apps can also vary from city to city.
This observation aligns with one participant's statement (P384),
criticizing that "the scenario earthquake is implausible for Bremen. [\dots] A more suitable example would be flooding."
For example, the Tier 3 app category weather apps should include tide forecasts for areas close to the sea, while areas in the mountains need rain and snow forecasts.
On the other hand, managing app stores inflicts a lot of effort on municipalities, and leaving distributed crisis app stores under the authority of app operators or network providers might scale better.

\paragraph{Towards Island-Ready Users}

Although our sample is limited to adult smartphone users from major German cities, previous research has identified different national risk cultures \cite{cornia_risk_2016}.
\textit{State-oriented} risk cultures emphasize the state's responsibility for crisis prevention and management.
\textit{Individual-oriented} risk cultures describe that citizens assume their duties of being informed, prepared, and aware concerning risks.
\textit{Fatalistic} risk cultures are characterized by low confidence in their respective individual and state problem-solving potentials.
While an appropriate level of both \designcriterion{citizen and state preparedness} and \designcriterion{response capabilities} is necessary, it seems essential to examine educational, political, and technological measures to prepare users and policymakers for the scenario of island connectivity.
This shapes \designcriterion{transparency of smartphone} \designcriterion{service prioritization} to mitigate potential frustration upon the unavailability of functionality and \designcriterion{trust into critical infrastructure} \designcriterion{resilience,} including the expected outage durations \cite{kaufhold_potentiale_2019}.

\subsection{Outlook and Future Work}
\label{ssec: discussion-future-work}

In this paper, we have introduced the concept of island connectivity in crisis-struck areas to the \ac{HCI} community.
We acknowledge that this concept is visionary, and future work needs to solve many challenges before its realization.

\paragraph{Future HCI research}
Our survey studied how adult smartphone users \new{(over 18 and under 70)} from major German cities prefer using smartphones in isolated cities.
\new{We decided to screen out other groups to align our study sample with our study subject.
Nevertheless, insights from groups under-represented in our study can complement our findings very well, and we explicitly encourage future surveys that build upon our findings by studying other samples, e.g., participants from rural areas or from other countries.}
Upon acceptance of this paper, we will publish a replication package to facilitate such future studies.
Our goal was to understand users' preferences about smartphone usage on islands with a focus on statistical and quantitative evaluation.
Future qualitative studies can help refine our quantitative and statistical findings.
Interviewing experts from app development, \acp{MNO}, and authorities can augment our user-centric findings with insights from other involved parties.
\new{Surveys closer to real-world crisis scenarios, e.g., conducted in an affected area during crisis response, can support or contradict our findings. Connecting HCI research with one of the few available crisis-related datasets \cite{alvarez_2018_conducting, kim_2018_social} can create valuable insights.}

\paragraph{Future 6G research}
The design implications discussed in \cref{ssec: discussion-implications} give a first glance at the challenges left to be solved by future work.
Most importantly, including distributed core networks in \ac{6G} is a paramount enabler of local connectivity in isolated areas.
Future research can support app developers in making their smartphone apps island-ready by identifying suitable communication patterns for a swift transition into island operation.

\section{Conclusion}
\label{sec: Conclusion}

The work reported in this paper makes two central contributions:
\new{Introducing the conceptual model of island connectivity and conducting the first survey ($N = \nfinal$) of user preferences regarding smartphone usage on islands with local-only connectivity.}
With the upcoming standardization of \ac{6G}, it is crucial to understand the user perspective of island connectivity to support operators, developers, and authorities in making user-centric design decisions while realizing island connectivity.

\new{We describe how \ac{6G} can facilitate island connectivity for crisis-struck areas and introduce the concept of island connectivity.}
We apply statistical and quantitative methods to learn that users prefer apps other than crisis apps for some crisis-specific use cases, e.g., messengers for communicating with other citizens and news apps to look up information (\researchquestion{RQ1}).
App demands differ greatly in isolated areas compared to everyday life (\researchquestion{RQ2}).
While news and crisis apps surge in demand, shopping and travel apps plummet.
These findings are based on our survey whose design we describe in detail, proactively considering biases and limitations.
We combine our survey's insights with applicable standards to develop a smartphone service typology (\researchquestion{RQ3}), organizing apps in tiers with descending criticality for isolated cities, distinguishing between apps crucial for crisis response and apps enabling crisis routine.
From this classification, we derive implications for mobile network operators, app developers, and authorities, facilitating user-centric design decisions in the realization process of island-ready 6G.

\begin{acks}
This work has been co-funded by the Federal Ministry of Education and Research of Germany in the projects Open6GHub (grant number: 16KISK014) and CYLENCE (grant number: 13N16636), and by the LOEWE initiative (Hesse, Germany) within the emergenCITY center [LOEWE/1/12/519/03/05.001(0016)/72].
We thank the participants of our pilot study and the anonymous reviewers for their helpful suggestions, which helped improve this work.
We thank Lea Holaus for the comics (Figures~\ref{fig:comic1} and \ref{fig:comic2}), Alexander Heinrich for supporting the qualitative data analysis, and Sebastian Perrig and Nicholas Scharowski for the helpful discussions on survey design.
\end{acks}

\section*{Availability}
We provide artifacts to verify, reproduce, and replicate our findings:
With this paper, we publish our survey dataset and evaluation scripts \cite{zenodo_2025_dataset} to verify and reproduce our results and the LimeSurvey survey structure to replicate our study.

\bibliographystyle{ACM-Reference-Format}



\appendix

\section{Questionnaire}
\label{sec: survey}

We translated the transcript of the questionnaire from German to English.
We did not provide the headlines and part descriptions during the survey.

\subsection{Part 1: Consent and Screening}
\surveyquestion{Age (Q1)}
"How old are you?"
\surveyansweroptions{"Under 18", "18-29", "30-39", "40-49", "50-59", "60-69", "70+"}
\surveyquestion{Gender (Q2)}
"Which gender do you identify with?"
\surveyansweroptions{"Female," "Male," "Non-Binary," "Own description" with a text field, "Prefer not to disclose"}
\surveyquestion{Education (Q3)}
"What is your highest educational qualification?"
\surveyansweroptions{"None (yet)," "Hauptschulabschluss," "Polytechnische Oberschule," "Mittlere Reife, Realschulabschluss," "(Fach-)Hochschulreife," "Bachelor," "Master," "Diplom,"  "Promotion," "Other qualification" with a textfield}
\surveyquestion{Income (Q4)}
"If you add up all the incomes of your household, Which of the following income groups does your monthly net household income fall into (i.e., household income after deducting taxes)?"
\surveyansweroptions{"under \SI{1000}{\EUR}", 
 "\SIrange{1000}{1999}{\EUR}", "\SIrange{2000}{2999}{\EUR}", "\SIrange{3000}{3999}{\EUR}", "\SIrange{4000}{4999}{\EUR}", "\SI{5000}{\EUR} and more", "No answer"}
\surveyquestion{Federal State (Q5)}
"In which federal state do you currently live?"
\surveyansweroptions{
"\acl{BW}",
"\acl{BY}",
"\acl{BE}",
"\acl{BB}",
"\acl{HB}",
"\acl{HH}",
"\acl{HE}",
"\acl{MV}",
"\acl{NI}",
"\acl{NW}",
"\acl{RP}",
"\acl{SL}",
"\acl{SN}",
"\acl{ST}",
"\acl{SH}",
"\acl{TH}"
}
\surveyquestion{Citizen and Smartphone Ownership (Q6)}
"Have you lived in a major German city (population at least \num{100,000}) for at least \SI{1}{year} since 2008?"
\surveyansweroptions{"Yes", "No"}
\surveyquestion{City Name (Q7)}
"What is the name of the German city where you have lived for at least one year since 2008?"
\surveyansweroptions{
  "Aachen",
  "Augsburg",
  "Bergisch Gladbach",
  "Berlin",
  "Bielefeld",
  "Bochum",
  "Bonn",
  "Bottrop",
  "Braunschweig",
  "Bremen",
  "Bremerhaven",
  "Chemnitz",
  "Darmstadt",
  "Dortmund",
  "Dresden",
  "Duisburg",
  "Düsseldorf",
  "Erfurt",
  "Erlangen",
  "Essen",
  "Frankfurt am Main",
  "Freiburg im Breisgau",
  "Fürth",
  "Gelsenkirchen,"
  "Göttingen",
  "Gütersloh",
  "Hagen",
  "Halle (Saale)",
  "Hamburg",
  "Hamm",
  "Hanau",
  "Hannover",
  "Heidelberg",
  "Heilbronn",
  "Herne",
  "Hildesheim",
  "Ingolstadt",
  "Jena",
  "Kaiserslautern",
  "Karlsruhe",
  "Kassel",
  "Kiel",
  "Koblenz",
  "Köln",
  "Krefeld",
  "Leipzig",
  "Leverkusen",
  "Lübeck",
  "Ludwigshafen am Rhein",
  "Magdeburg",
  "Mainz",
  "Mannheim",
  "Moers",
  "Mönchen-Gladbach",
  "Mülheim an der Ruhr",
  "München",
  "Münster",
  "Neuss",
  "Nürnberg",
  "Oberhausen",
  "Offenbach am Main",
  "Oldenburg (Oldb)",
  "Osnabrück,"
  "Paderborn",
  "Pforzheim",
  "Potsdam",
  "Recklinghausen",
  "Regensburg",
  "Remscheid",
  "Reutlingen,"
  "Rostock",
  "Saarbrücken",
  "Salzgitter",
  "Siegen",
  "Solingen",
  "Stuttgart",
  "Trier",
  "Ulm",
  "Wiesbaden",
  "Wolfsburg",
  "Wuppertal",
  "Würzburg"
}
\surveyquestion{City Duration (Q8)}
"How many years have you lived in this city since 2008?"
\surveyansweroptions{Numerical input field}
\surveyquestion{City Smartphone Usage Confirmation (Q9)}
"Have you owned and used a smartphone during this time?"
\surveyansweroptions{"Yes", "No"}

\subsection{Part 2: Everyday Smartphone Usage}
\surveyquestion{Daily Smartphone Screentime (Q10)}
"Please estimate your average daily smartphone screen time in hours."
\surveyhint{By screen time, we mean the time you actively use your smartphone. For example, Apple smartphones offer these values via the 'Screen Time' function. On newer Android smartphones, this function is called 'Digital Wellbeing.'}
\surveyansweroptions{Numerical input field}
\surveyquestion{App Usage in Everyday Life (Q11)}
"Please sort the following app categories according to the frequency with which you use apps from these categories. Sort in descending order, i.e., the most frequently used app category should come first, and the least used app category should be last. If you never use certain app categories, you do not have to include them and can ignore them. However, you must specify at least one category."
\surveyhint{The apps in brackets only suggest the respective categories. The mouse can move the elements between the left and right lists. A double-click moves an element to the other list.}
\surveyansweroptions{positional voting of \texttt{app\_categories}}

\subsection{Part 3: Crisis Smartphone Usage}

\surveyquestion{Crisis Introduction (Comic)}
"Below is a short comic board that introduces you to the scenario of this survey. When we mention major cities, please think of \texttt{the\_city}."
\textit{The comic board as depicted in Figure~\ref{fig:comic1}.}
\surveyquestion{App Demand for Crisis-Specific Use Cases (Q12)}
"In the previous comic, an earthquake was introduced as an example of a crisis.
For the following crises, please indicate which apps you would use.
The categories are the same as in Q11. Multiple answers are possible."
(U1) "You want to receive warnings about the crisis. Which apps would you use?"
\surveyansweroptions{\texttt{app\_categories}, "Other" with a text field, "No app," "No answer"}
(U2) "You want to receive tips on how to stay out of danger in the event of a crisis. Which apps would you use?"
\surveyansweroptions{\texttt{app\_categories}, "Other" with a text field, "No app," "No answer"}
(U3) "You want to contact an emergency service (but calling 112 is not possible). Which apps would you use?"
\surveyansweroptions{\texttt{app\_categories}, "Other" with a text field, "No app," "No answer"}
(U4) "You want to search for information about the crisis. Which apps would you use?"
\surveyansweroptions{\texttt{app\_categories}, "Other" with a text field, "No app," "No answer"}
(U5) "You want to contact and communicate with other citizens. Which apps would you use?"
\surveyansweroptions{\texttt{app\_categories}, "Other" with a text field, "No app," "No answer"}
(U6) "You want to navigate in \texttt{the\_city}. Which apps would you use?"
\surveyansweroptions{\texttt{app\_categories}, "Other" with a text field, "No app," "No answer"}
(U7) "You want to be distracted from the crisis. Which apps would you use?"
\surveyansweroptions{\texttt{app\_categories}, "Other" with a text field, "No app," "No answer"}

\subsection{Part 4: Crisis Smartphone Interaction with Island Connectivity}

\surveyquestion{City Isolation Introduction (Comic)}
"The following comic builds on the previous scenario and expands on it, continuing the earthquake scenario in \texttt{the\_city}."
\textit{The comicboard as depicted in Figure~\ref{fig:comic2}.}
\surveyquestion{City Isolation Introduction (Video)}
\label{sssec: isolated-city-video}
"Please start the video if it does not start automatically. If this does not work either, there is a description of the video below the video."
\surveyquestion{Video Transcript}
"First, you want to send a message to a friend outside the city using a messenger app. The message is usually sent via the Internet. However, the earthquake disconnected the city from the Internet, so the message could not reach its destination.
After the recovery measures taken by the mobile network operator, all apps on your smartphone are available for local use. Therefore, you can continue using your already installed apps to a limited extent, e.g., you can only navigate inside the city with your navigation apps, only read the local newspaper app but not global newspaper apps, or only communicate with citizens inside the city with your messenger apps, but not with people outside the city."
\surveyquestion{Video Description}
"You can see how a chat message, symbolized by a letter, is to be sent to another city. However, the message cannot be sent because your city is disconnected from the Internet.
You can then see that the mobile operator enables the use of your smartphone apps locally. This means, for example, that you can only use your navigation apps to navigate within the city, that you can only read your local newspaper app but not a global news app, or that you can only use your messenger apps to communicate with people within the city but not with people outside it."
\surveyquestion{App Usage in Isolated Cities (Q13)}
"The consequences of the earthquake have now lasted for several days.
For this question, please assume that your apps work locally in the city, as the video explains.
Apps from all categories are available in this way.
Such local recovery measures are visions for the future that have not yet been implemented.
To answer this question, please imagine that apps on your smartphone work locally, as shown in the comic and video.
The categories are the same as for Q11 and Q12.
Please sort the following app categories according to the frequency you use apps from these categories.
Sort in descending order, i.e., the most frequently used app category should come first, and the least used app category should be last.
If you never use certain app categories, you do not have to include them and can ignore them.
However, you must specify at least one category."
\surveyhint{The apps in brackets only suggest the respective categories. The mouse can move the elements between the left and right lists. A double-click moves an element to the other list.}
\surveyansweroptions{positional voting of \texttt{app\_categories}}

\subsection{Part 5: Control Variables and Data Quality}

"The following questions are independent of the previous scenario and relate only to you as a person."
\surveyquestion{Crisis Experience (Q14)}
"Please indicate which of the following crises you have experienced."
\surveyhint{We are not counting the COVID-19 pandemic as a crisis here.}
\surveyansweroptions{"Earthquake," "Flooding," "Storm," 'Wildfire,' "War," "None," "Not sure" with a textfield}
\surveyquestion{Internet Outage Experience (Q15)}
"Please indicate which of the following Internet outages you have experienced."
\surveyhint{Fixed network refers to your Internet access at home, e.g., via Wi-Fi.}
"No Internet via fixed network but via cellular network."
\surveyansweroptions{"Never," "up to 1h", "up to 6h", "up to 12h", "up to 24h", "more than 24h", "Not sure"}
"No Internet via cellular network but via fixed network."
\surveyansweroptions{"Never," "up to 1h", "up to 6h", "up to 12h", "up to 24h", "more than 24h", "Not sure"}
"No Internet via fixed network nor cellular network."
\surveyansweroptions{"Never," "up to 1h", "up to 6h", "up to 12h", "up to 24h", "more than 24h", "Not sure"}
\surveyquestion{Emergency Forces Experience (Q16)}
"Are you currently working as an emergency worker, e.g., for the fire department, rescue service, police, or THW?"
\surveyansweroptions{"Yes," "No," "No answer," "Other" with a textfield}
\surveyquestion{ATI Scale with IRI (Q17)}
"The following is about your interaction with technical systems.
By 'technical systems," we mean apps and other software applications as well as complete digital devices (e.g., cell phone, computer, TV, car navigation).
Please indicate your level of agreement with the following statements."
"I like to occupy myself in greater detail with technical systems."
\surveyansweroptions{"Completely disagree," "Largely disagree," "Slightly disagree," "Slightly agree," "Largely agree," "Completely agree"}
"I like testing the functions of new technical systems."
\surveyansweroptions{"Completely disagree," "Largely disagree," "Slightly disagree," "Slightly agree," "Largely agree," "Completely agree"}
"I predominantly deal with technical systems because I have to."
\surveyansweroptions{"Completely disagree," "Largely disagree," "Slightly disagree," "Slightly agree," "Largely agree," "Completely agree"}
"When I have a new technical system in front of me, I try it extensively."
\surveyansweroptions{"Completely disagree," "Largely disagree," "Slightly disagree," "Slightly agree," "Largely agree," "Completely agree"}
"I enjoy spending time becoming acquainted with a new technical system."
\surveyansweroptions{"Completely disagree," "Largely disagree," "Slightly disagree," "Slightly agree," "Largely agree," "Completely agree"}
"It is enough for me that a technical system works; I don't care how and why."
\surveyansweroptions{"Completely disagree," "Largely disagree," "Slightly disagree," "Slightly agree," "Largely agree," "Completely agree"}
"Please indicate 'Completely disagree' here."
\surveyansweroptions{"Completely disagree," "Largely disagree," "Slightly disagree," "Slightly agree," "Largely agree," "Completely agree"}
"I try to understand how a technical system exactly works."
\surveyansweroptions{"Completely disagree," "Largely disagree," "Slightly disagree," "Slightly agree," "Largely agree," "Completely agree"}
"It is enough for me to know the basic functions of a technical system."
\surveyansweroptions{"Completely disagree," "Largely disagree," "Slightly disagree," "Slightly agree," "Largely agree," "Completely agree"}
"I try to make full use of the capabilities of a technical system."
\surveyansweroptions{"Completely disagree," "Largely disagree," "Slightly disagree," "Slightly agree," "Largely agree," "Completely agree"}
\surveyquestion{Technical Issues Textfield (Q18)}
"Did you have any technical problems answering the questions?"
\surveyansweroptions{Text field}
\surveyquestion{Additional Comments Textfield (Q19)}
"If you have anything else you want to tell us, please use the text field below."
\surveyansweroptions{Text field}
\surveyquestion{SRSI UseMe (Q20)}
"Please be honest: This survey is important to our research. Should we use your answers for our studies?"
\surveyhint{Do not worry; the answer to this question is only for our data analysis. We will not pass them on to GapFish or third parties; they will not affect your payment.}
\surveyansweroptions{"Yes", "No"}

\begin{table*}[t]
  \caption{
      \captionheadline{Rank Distributions by Variable and Group}
      For each variable, this table compares the median rank $M_g(a)$ that app category $a$ has scored in group $g$.
      There is a row per group $g$, e.g., younger participants or participants with medium technology affinity, and a column per app category $a$, e.g., messengers or news apps.
      Our participants did not have to rank all app categories, and those that they left unranked are consequently not included in the median calculation.
      Note that our overall rank calculation in Table~\ref{tab:rq2} uses the Borda count \cite{mclean_classics_1995}, which accounts for unranked apps during the calculation (e.g., for crisis apps, which many participants did not include).}
  \Description{
      25 columns: Variable, Category and the 23 app categories ranked like in Table 2b: messengers, news, crisis, social media, weather, web browsers, navigation, entertainment, music, utilities, reference, finance, food \& drink, games, productivity, books, health \& fitness, medical, shopping, education, travel, creativity, sports.
      There are rows for groups of the following variables: Age, Gender, Education, Income, Federal State, City Size, Residence Time, Screen Time, Crisis Experience, Outage Experience, First Responder, and Technology Affinity.
  }
\label{tab:rankdist-all}
\small
\setlength{\tabcolsep}{2pt}
\begin{tabular}{ll rrr rrrrr rrrrr rrrrr rrrrr}
  
  \textbf{Variable}  & \textbf{Category} &
  \rotatebox{90}{\textbf{Messengers}} &
  \rotatebox{90}{\textbf{News}} &
  \rotatebox{90}{\textbf{Crisis}} &
  \rotatebox{90}{\textbf{Social Media}} &
  \rotatebox{90}{\textbf{Weather}} &
  \rotatebox{90}{\textbf{Web Browsers}} &
  \rotatebox{90}{\textbf{Navigation}} &
  \rotatebox{90}{\textbf{Entertainment}} &
  \rotatebox{90}{\textbf{Music}} &
  \rotatebox{90}{\textbf{Utilities}} &
  \rotatebox{90}{\textbf{Reference}} &
  \rotatebox{90}{\textbf{Finance}} &
  \rotatebox{90}{\textbf{Food \& Drink}} &
  \rotatebox{90}{\textbf{Games}} &
  \rotatebox{90}{\textbf{Productivity}} &
  \rotatebox{90}{\textbf{Books}} &
  \rotatebox{90}{\textbf{Health \& Fitness}} &
  \rotatebox{90}{\textbf{Medical}} &
  \rotatebox{90}{\textbf{Shopping}} &
  \rotatebox{90}{\textbf{Education}} &
  \rotatebox{90}{\textbf{Travel}} &
  \rotatebox{90}{\textbf{Creativity}} &
  \rotatebox{90}{\textbf{Sports}} \\
  
  \hline 
  
\cellcolor{aa}                      & \cellcolor{aa}Younger       & \cellcolor{aa}      2 & \cellcolor{aa}3 & \cellcolor{aa}2   &      \cellcolor{aa}   4 & \cellcolor{aa}6   &  \cellcolor{aa}     6   & \cellcolor{aa}      6 &  \cellcolor{aa}      7   & \cellcolor{aa}5   & \cellcolor{aa}  6   & \cellcolor{aa}  7   & \cellcolor{aa}7   &    \cellcolor{aa}   7   & \cellcolor{aa}10   & \cellcolor{aa}     10   & \cellcolor{aa}6   & \cellcolor{aa}         8   & \cellcolor{aa}11   &  \cellcolor{aa}11   & \cellcolor{aa}  4   & \cellcolor{aa}11   &  \cellcolor{aa}  12   & \cellcolor{aa}12.5 \\
  \rccc{aa}{aa}                           & \cellcolor{aa}Middle-Aged   & \cellcolor{aa}      2 & \cellcolor{aa}3 & \cellcolor{aa}1   &  \cellcolor{aa}       4 & \cellcolor{aa}6   & \cellcolor{aa}     5   & \cellcolor{aa}      5 & \cellcolor{aa}       8   & \cellcolor{aa}7   & \cellcolor{aa}   6   & \cellcolor{aa}   7   & \cellcolor{aa} 8   & \cellcolor{aa}     8   & \cellcolor{aa}9   &    \cellcolor{aa}  10   & \cellcolor{aa}7.5 &    \cellcolor{aa}       9.5 &  \cellcolor{aa}8   &   \cellcolor{aa}11   & \cellcolor{aa}  8   & \cellcolor{aa}14   &    \cellcolor{aa}12   &  \cellcolor{aa}14 \\
  \rccc{aa}{aa} \multirow{-3}*{Age}  & \cellcolor{aa}Older         &     \cellcolor{aa}   2 & \cellcolor{aa}3 &   \cellcolor{aa}1   &   \cellcolor{aa}       4 & \cellcolor{aa} 5   &  \cellcolor{aa}      5   & \cellcolor{aa}       4 & \cellcolor{aa}        8   & \cellcolor{aa}12   &  \cellcolor{aa}   6   & \cellcolor{aa}   10  \cellcolor{aa}&    \cellcolor{aa}9   &       \cellcolor{aa}10   & \cellcolor{aa}11   &   \cellcolor{aa}     9.5 & \cellcolor{aa}5.5 & \cellcolor{aa}          13   & \cellcolor{aa}  7   & \cellcolor{aa} 14   & \cellcolor{aa}   9.5 & \cellcolor{aa}12   &  \cellcolor{aa}   16.5 & \cellcolor{aa}12 \\
  
  \hline 
  
  \cellcolor{ba}                         & \cellcolor{ba}Men          &  \cellcolor{ba} 2 &  \cellcolor{ba}3 &   \cellcolor{ba}2   &     \cellcolor{ba}4 &\cellcolor{ba}6   & \cellcolor{ba} 5   &\cellcolor{ba}   5 & \cellcolor{ba}  8   & \cellcolor{ba}6   &   \cellcolor{ba}5   &  \cellcolor{ba}8   & \cellcolor{ba}7   &   \cellcolor{ba}8   & \cellcolor{ba}10   & \cellcolor{ba}10   &  \cellcolor{ba}8   &      \cellcolor{ba} 8   & \cellcolor{ba}8   & \cellcolor{ba}12   &  \cellcolor{ba}6.5 & \cellcolor{ba}13   &    \cellcolor{ba}13   &  \cellcolor{ba}12 \\
  \rccc{ba}{ba}                               &\cellcolor{ba}Women   &  \cellcolor{ba} 2 & \cellcolor{ba}3 &  \cellcolor{ba}2   &    \cellcolor{ba} 4 & \cellcolor{ba}6   & \cellcolor{ba} 6   & \cellcolor{ba} 6 & \cellcolor{ba}  7   &  \cellcolor{ba}6   &  \cellcolor{ba}7   & \cellcolor{ba}6   & \cellcolor{ba}8   &  \cellcolor{ba} 8   & \cellcolor{ba}10   &     \cellcolor{ba}9.5 & \cellcolor{ba}6   &   \cellcolor{ba}   10   & \cellcolor{ba}10   &  \cellcolor{ba}10   &  \cellcolor{ba}7   & \cellcolor{ba}11   &  \cellcolor{ba}12   & \cellcolor{ba}16 \\
  \rccc{ba}{ba} \multirow{-3}*{Gender}   &\cellcolor{ba}Non-Binary        &   \cellcolor{ba}3 &  \cellcolor{ba}4 &  \cellcolor{ba}4   &     \cellcolor{ba}4 &  \cellcolor{ba}9   &    \cellcolor{ba}3   &  \cellcolor{ba} 8 &  \cellcolor{ba}  4.5 & \cellcolor{ba}8   &    \cellcolor{ba}&     \cellcolor{ba}5   &   \cellcolor{ba}4   &      \cellcolor{ba}6   & \cellcolor{ba}3.5 &      \cellcolor{ba} &  \cellcolor{ba}2   &   \cellcolor{ba}    1   & \cellcolor{ba}2   &  \cellcolor{ba}9.5 &    \cellcolor{ba}& \cellcolor{ba}&   \cellcolor{ba}  &\cellcolor{ba}\\
  
  \hline 
  \rccc{aa}{aa}
\cellcolor{aa}                                           &  \cellcolor{aa}Low    &     \cellcolor{aa}  2 & \cellcolor{aa}3 & \cellcolor{aa}1   &   \cellcolor{aa}      4 & \cellcolor{aa}5   &    \cellcolor{aa}   5.5 &   \cellcolor{aa}    4 &  \cellcolor{aa}      6   & \cellcolor{aa}6   & \cellcolor{aa}  5   & \cellcolor{aa}  5   & \cellcolor{aa}14   &    \cellcolor{aa}   9   &  \cellcolor{aa}8.5 &    \cellcolor{aa}  10.5 &  \cellcolor{aa}6   &     \cellcolor{aa}    10   & \cellcolor{aa}5.5 &   \cellcolor{aa}8   &    \cellcolor{aa}11   & \cellcolor{aa}11.5 &     \cellcolor{aa}16   &  \cellcolor{aa}13 \\
  \rccc{aa}{aa}  
\cellcolor{aa}                                         & \cellcolor{aa}Medium       & \cellcolor{aa}     2 &  \cellcolor{aa}3 &   \cellcolor{aa}2   &   \cellcolor{aa}      4 & \cellcolor{aa}6   &  \cellcolor{aa}     5   &  \cellcolor{aa}     5 & \cellcolor{aa}      8   & \cellcolor{aa}6   & \cellcolor{aa}  6   & \cellcolor{aa}  7   & \cellcolor{aa}8   & \cellcolor{aa}     8   & \cellcolor{aa}10   &  \cellcolor{aa}    11   & \cellcolor{aa}5   &      \cellcolor{aa}     9   &  \cellcolor{aa}10   &  \cellcolor{aa}12   & \cellcolor{aa}  6   & \cellcolor{aa}13   &   \cellcolor{aa} 13   &  \cellcolor{aa}13 \\
  \rccc{aa}{aa} 
  \multirow{-3}*{Education}   & \cellcolor{aa}High      &   \cellcolor{aa}    2 & \cellcolor{aa}3 &  \cellcolor{aa}2   &       \cellcolor{aa}  4 &    \cellcolor{aa}6   &       \cellcolor{aa}6   &      \cellcolor{aa} 5 &   \cellcolor{aa}     7   & \cellcolor{aa}5   &  \cellcolor{aa}  7   &   \cellcolor{aa} 8   &  \cellcolor{aa}7   &  \cellcolor{aa}     8   & \cellcolor{aa}10   &  \cellcolor{aa}     9   &  \cellcolor{aa}8   &     \cellcolor{aa}      9   & \cellcolor{aa}10   &   \cellcolor{aa}11   &    \cellcolor{aa}7   &  \cellcolor{aa}11   &   \cellcolor{aa} 12.5 &  \cellcolor{aa}14 \\
  
  \hline 
  
  \rccc{ba}{ba} 
\cellcolor{ba}
                                & \cellcolor{ba}Low      &   \cellcolor{ba}2 & \cellcolor{ba}3 & \cellcolor{ba}1   &    \cellcolor{ba} 4 & \cellcolor{ba}5   &  \cellcolor{ba} 5   & \cellcolor{ba} 5 &  \cellcolor{ba}  7   & \cellcolor{ba}6   &  \cellcolor{ba}6   &  \cellcolor{ba}6   &\cellcolor{ba}11   &    \cellcolor{ba}9.5 &  \cellcolor{ba}8.5 &        \cellcolor{ba}8   & \cellcolor{ba}6   &     \cellcolor{ba} 10   &\cellcolor{ba}7   &   \cellcolor{ba}11   &   \cellcolor{ba}5   &  \cellcolor{ba}11   &     \cellcolor{ba}13   & \cellcolor{ba}14.5 \\
  \rccc{ba}{ba}  
   \cellcolor{ba}                    & \cellcolor{ba}Medium         &   \cellcolor{ba}2 &  \cellcolor{ba}3 &  \cellcolor{ba}2   &    \cellcolor{ba} 4 & \cellcolor{ba}6   &  \cellcolor{ba} 5   & \cellcolor{ba} 5 &   \cellcolor{ba} 7.5 &\cellcolor{ba}6   & \cellcolor{ba}6   & \cellcolor{ba}7   &\cellcolor{ba}7   &    \cellcolor{ba}8   &  \cellcolor{ba}9.5 &    \cellcolor{ba}10   &  \cellcolor{ba}6   &  \cellcolor{ba}    10   &  \cellcolor{ba}10.5 &   \cellcolor{ba}10   &   \cellcolor{ba}7.5 &  \cellcolor{ba}11   &   \cellcolor{ba}12   &  \cellcolor{ba}12 \\
\rccc{ba}{ba} 
\cellcolor{ba}
\multirow{-3}*{Income}   & \cellcolor{ba}High        &  \cellcolor{ba} 2 &  \cellcolor{ba}3 &   \cellcolor{ba}2   &  \cellcolor{ba}   4 & \cellcolor{ba}6   & \cellcolor{ba} 6   & \cellcolor{ba} 6 &   \cellcolor{ba} 8   &\cellcolor{ba}7   &  \cellcolor{ba}7   &  \cellcolor{ba}8   &  \cellcolor{ba}7   &    \cellcolor{ba}7   & \cellcolor{ba}11   &     \cellcolor{ba}10   &  \cellcolor{ba}7   &    \cellcolor{ba}   8   & \cellcolor{ba}9   &   \cellcolor{ba}13.5 &    \cellcolor{ba}6   &  \cellcolor{ba}14   &    \cellcolor{ba}14   &  \cellcolor{ba}14.5 \\

  \hline 
  
\rccc{aa}{aa}
\cellcolor{aa}
                                   & \cellcolor{aa}Medium      &   \cellcolor{aa}    2 &  \cellcolor{aa}3 &   \cellcolor{aa}1.5 &      \cellcolor{aa}   4 & \cellcolor{aa}5.5 &     \cellcolor{aa}  5   &  \cellcolor{aa}     5 &  \cellcolor{aa}      6   & \cellcolor{aa}5.5 &     \cellcolor{aa}6   & \cellcolor{aa}  7   & \cellcolor{aa}7   &  \cellcolor{aa}     6   & \cellcolor{aa}9   &   \cellcolor{aa}    9   & \cellcolor{aa}9   &    \cellcolor{aa}      11   &   \cellcolor{aa}10.5 &    \cellcolor{aa}10   &  \cellcolor{aa}  8.5 &  \cellcolor{aa}10.5 &     \cellcolor{aa}13.5 &  \cellcolor{aa}9 \\
  \rccc{aa}{aa} 
 \cellcolor{aa}                                & \cellcolor{aa}Smaller       &  \cellcolor{aa}     2 & \cellcolor{aa}3 &   \cellcolor{aa}1   &    \cellcolor{aa}     4 & \cellcolor{aa}6.5 &   \cellcolor{aa}    6   & \cellcolor{aa}     5 &    \cellcolor{aa}   10   & \cellcolor{aa}7   &   \cellcolor{aa} 5.5 &   \cellcolor{aa} 5.5 &  \cellcolor{aa}9.5 &     \cellcolor{aa}  8   & \cellcolor{aa}11   &  \cellcolor{aa}    11.5 &  \cellcolor{aa}5   &      \cellcolor{aa}     7.5 &   \cellcolor{aa}8.5 &    \cellcolor{aa}12   &  \cellcolor{aa}  5   &  \cellcolor{aa}15   &  \cellcolor{aa}  14   &  \cellcolor{aa}16 \\
 \rccc{aa}{aa}
\cellcolor{aa}                                 & \cellcolor{aa}Larger   &   \cellcolor{aa}    2 & \cellcolor{aa}3 &  \cellcolor{aa}2   &    \cellcolor{aa}     4 & \cellcolor{aa}6   & \cellcolor{aa}     5.5 &  \cellcolor{aa}     6 &  \cellcolor{aa}      8   & \cellcolor{aa}8   & \cellcolor{aa}  6   & \cellcolor{aa}  9   &  \cellcolor{aa}8   &  \cellcolor{aa}     9   & \cellcolor{aa}9.5 &  \cellcolor{aa}    10   & \cellcolor{aa}12   &     \cellcolor{aa}      9   & \cellcolor{aa}9   &  \cellcolor{aa}12   & \cellcolor{aa}  9   & \cellcolor{aa}12.5 & \cellcolor{aa}  10   &  \cellcolor{aa}12 \\
  \rccc{aa}{aa} 
  \cellcolor{aa}
  \multirow{-4}*{City Size}         &  \cellcolor{aa}Metropolis       &         \cellcolor{aa}2 &   \cellcolor{aa}3 &   \cellcolor{aa}1   &    \cellcolor{aa}     4 & \cellcolor{aa}5   &  \cellcolor{aa}     5   &   \cellcolor{aa}    5 &    \cellcolor{aa}    7   &  \cellcolor{aa}5   & \cellcolor{aa}   6   & \cellcolor{aa}   6   & \cellcolor{aa}7.5 & \cellcolor{aa}     7.5 & \cellcolor{aa}9   &      \cellcolor{aa} 8   & \cellcolor{aa}6   &    \cellcolor{aa}      10   & \cellcolor{aa}10   & \cellcolor{aa}  8.5 & \cellcolor{aa}   5   &  \cellcolor{aa}11   &     \cellcolor{aa}15   &  \cellcolor{aa}15 \\
  
  \hline 
  
\rccc{ba}{ba}   
\cellcolor{ba}
                                    & \cellcolor{ba}Short    & \cellcolor{ba} 2 & \cellcolor{ba}2 &  \cellcolor{ba}2   &  \cellcolor{ba}   5 &  \cellcolor{ba}7   &   \cellcolor{ba}4   &  \cellcolor{ba} 5 &    \cellcolor{ba}7   & \cellcolor{ba}5.5 &   \cellcolor{ba}8   &  \cellcolor{ba}7   &  \cellcolor{ba}10   &  \cellcolor{ba} 8.5 & \cellcolor{ba}10   & \cellcolor{ba} 8   & \cellcolor{ba}7.5 &   \cellcolor{ba}   16.5 &  \cellcolor{ba}15   &   \cellcolor{ba}12   & \cellcolor{ba}4   &  \cellcolor{ba}10   &    \cellcolor{ba}14   &  \cellcolor{ba}12 \\
  \rccc{ba}{ba} 
\cellcolor{ba}                                & \cellcolor{ba}Medium     &    \cellcolor{ba}2 &\cellcolor{ba}3 & \cellcolor{ba}2   &   \cellcolor{ba}  5 &\cellcolor{ba}7.5 &  \cellcolor{ba} 6.5 &   \cellcolor{ba}5 &     \cellcolor{ba}8   &\cellcolor{ba}5   &   \cellcolor{ba}6   & \cellcolor{ba}7   &\cellcolor{ba}9   &  \cellcolor{ba} 8   &\cellcolor{ba}11   &  \cellcolor{ba} 7.5 & \cellcolor{ba}7   &      \cellcolor{ba} 8   &\cellcolor{ba}9   &    \cellcolor{ba}11.5 &   \cellcolor{ba}7   & \cellcolor{ba}11.5 &  \cellcolor{ba}13   &\cellcolor{ba}13 \\
\rccc{ba}{ba}
\cellcolor{ba}
   \multirow{-3}*{Residence Time}   & \cellcolor{ba}Long      &     \cellcolor{ba}2 & \cellcolor{ba}3 &  \cellcolor{ba}1   &      \cellcolor{ba}4 &  \cellcolor{ba}6   &  \cellcolor{ba} 5   &  \cellcolor{ba} 5 &   \cellcolor{ba} 7   &\cellcolor{ba}7   & \cellcolor{ba}6   & \cellcolor{ba}7   &   \cellcolor{ba}7   &  \cellcolor{ba} 8   & \cellcolor{ba}9   &  \cellcolor{ba}10   & \cellcolor{ba}6   &        \cellcolor{ba}9   &  \cellcolor{ba}9   &  \cellcolor{ba}11   &   \cellcolor{ba}7   &  \cellcolor{ba}13   &     \cellcolor{ba}13   & \cellcolor{ba}13 \\
  
  \hline 
  
\rccc{aa}{aa}     
\cellcolor{aa}                            & \cellcolor{aa}Medium   &   \cellcolor{aa}    2 &  \cellcolor{aa}3 & \cellcolor{aa}1   &     \cellcolor{aa}    4 & \cellcolor{aa}5   &  \cellcolor{aa}     5   &    \cellcolor{aa}   5 &      \cellcolor{aa}  7   & \cellcolor{aa}7   & \cellcolor{aa}  6   &  \cellcolor{aa}  7   &  \cellcolor{aa}8   &    \cellcolor{aa}   9   & \cellcolor{aa}11.5 &       \cellcolor{aa}10   & \cellcolor{aa}5   &      \cellcolor{aa}     9.5 &  \cellcolor{aa}8   & \cellcolor{aa}14   &      \cellcolor{aa}6.5 & \cellcolor{aa}14.5 &    \cellcolor{aa}13   &  \cellcolor{aa}15 \\
      \rccc{aa}{aa} 
\cellcolor{aa}    & \cellcolor{aa}High     &   \cellcolor{aa}    2 &  \cellcolor{aa}3 &  \cellcolor{aa}2   &       \cellcolor{aa}  4 &   \cellcolor{aa}6   &     \cellcolor{aa}  5   &      \cellcolor{aa} 5 &   \cellcolor{aa}     8   & \cellcolor{aa}5   & \cellcolor{aa}   6   &  \cellcolor{aa}  7   & \cellcolor{aa}7   &    \cellcolor{aa}   7   & \cellcolor{aa}10   &  \cellcolor{aa}     9   & \cellcolor{aa}7   &      \cellcolor{aa}     9.5 & \cellcolor{aa}10   &   \cellcolor{aa}10   &  \cellcolor{aa}  8   & \cellcolor{aa}11   &  \cellcolor{aa}  13.5 &  \cellcolor{aa}12 \\
 \rccc{aa}{aa} 
 \cellcolor{aa}
        \multirow{-3}*{Screen Time}                                   & \cellcolor{aa}Low      &  \cellcolor{aa}     2 & \cellcolor{aa}3 &   \cellcolor{aa}2   &        \cellcolor{aa} 4 &  \cellcolor{aa}6   &     \cellcolor{aa}  6   &   \cellcolor{aa}    6 &         \cellcolor{aa}8   & \cellcolor{aa}6   &    \cellcolor{aa}5.5 &    \cellcolor{aa}7   & \cellcolor{aa}8   &  \cellcolor{aa}     8   & \cellcolor{aa}9   &    \cellcolor{aa}  10.5 & \cellcolor{aa}7   & \cellcolor{aa}         9   & \cellcolor{aa}10   &    \cellcolor{aa}10   & \cellcolor{aa}  5   &  \cellcolor{aa}12   &      \cellcolor{aa}11.5 & \cellcolor{aa}13 \\
  
  \hline 
 \rccc{ba}{ba}      
 \cellcolor{ba}                                & \cellcolor{ba}Yes      &\cellcolor{ba}2          &\cellcolor{ba}3    &\cellcolor{ba}3      &\cellcolor{ba}4            &\cellcolor{ba}6       &\cellcolor{ba}5            &\cellcolor{ba}5          &\cellcolor{ba}11            &\cellcolor{ba}6     &\cellcolor{ba}5.5       &\cellcolor{ba}13        &\cellcolor{ba}7       &\cellcolor{ba}8            &\cellcolor{ba}19.5  &\cellcolor{ba}11           &\cellcolor{ba}7.5   &\cellcolor{ba}3                &\cellcolor{ba}15      &\cellcolor{ba}12       &\cellcolor{ba}5.5       &\cellcolor{ba}15.5   &\cellcolor{ba}16.5       &\cellcolor{ba}18 \\
\rccc{ba}{ba}
\cellcolor{ba}
     \multirow{-2}*{Crisis Experience}   &\cellcolor{ba}No       &\cellcolor{ba}2          &\cellcolor{ba}3    &\cellcolor{ba}1      &\cellcolor{ba}4            &\cellcolor{ba}6       &\cellcolor{ba}5            &\cellcolor{ba}5          &\cellcolor{ba}7             &\cellcolor{ba}6     &\cellcolor{ba}6         &\cellcolor{ba}7         &\cellcolor{ba}8       &\cellcolor{ba}8            &\cellcolor{ba}10    &\cellcolor{ba}10           &\cellcolor{ba}6     &\cellcolor{ba}10               &\cellcolor{ba}10      &\cellcolor{ba}11       &\cellcolor{ba}7         &\cellcolor{ba}12     &\cellcolor{ba}13         &\cellcolor{ba}13 \\
  
  \hline 
  
\rccc{aa}{aa} 
\cellcolor{aa} 
                                       & \cellcolor{aa}L. Out.: Yes       &    \cellcolor{aa}     2 &   \cellcolor{aa}3 &  \cellcolor{aa} 2   &        \cellcolor{aa}   4 & \cellcolor{aa}  5   &     \cellcolor{aa}    5   &     \cellcolor{aa}    5 &    \cellcolor{aa}      8   &  \cellcolor{aa}4   &  \cellcolor{aa}    4   &  \cellcolor{aa}   7.5 &  \cellcolor{aa}  6.5 &       \cellcolor{aa}  6   & \cellcolor{aa}8   &        \cellcolor{aa}10   & \cellcolor{aa}4.5 &     \cellcolor{aa}       10   &  \cellcolor{aa}  9   & \cellcolor{aa}   11   &  \cellcolor{aa}    6   & \cellcolor{aa} 11   & \cellcolor{aa}    11   & \cellcolor{aa}13 \\
  \rccc{aa}{aa}
\cellcolor{aa}
                                        & \cellcolor{aa}L. Out.: No       &    \cellcolor{aa}     2 &  \cellcolor{aa} 3 &  \cellcolor{aa} 1.5 &      \cellcolor{aa}     4 &   \cellcolor{aa} 6   &  \cellcolor{aa}       6   &  \cellcolor{aa}       5 &      \cellcolor{aa}    7   & \cellcolor{aa}6   &  \cellcolor{aa}    6   &     \cellcolor{aa} 7   &  \cellcolor{aa}  8   &     \cellcolor{aa}    8   & \cellcolor{aa}10   &    \cellcolor{aa}    10   & \cellcolor{aa} 7   &        \cellcolor{aa}     9   & \cellcolor{aa} 10   & \cellcolor{aa}  11   &    \cellcolor{aa}  7   & \cellcolor{aa} 12   &  \cellcolor{aa}    13   & \cellcolor{aa}13 \\
    \rccc{aa}{aa} 
    \cellcolor{aa}
                                       & \cellcolor{aa}C. Out.: Yes      &    \cellcolor{aa}     2 &  \cellcolor{aa} 3 & \cellcolor{aa} 2   &         \cellcolor{aa}  5 & \cellcolor{aa}   6   &   \cellcolor{aa}      6   &  \cellcolor{aa}       6 &      \cellcolor{aa}   11   & \cellcolor{aa} 6   &   \cellcolor{aa}   5.5 &   \cellcolor{aa}   8   &  \cellcolor{aa}  7   &   \cellcolor{aa}      8.5 & \cellcolor{aa}10.5 & \cellcolor{aa}      12   & \cellcolor{aa}4   &       \cellcolor{aa}      8   & \cellcolor{aa}  9   &  \cellcolor{aa}  11   &   \cellcolor{aa}   3.5 & \cellcolor{aa}14   & \cellcolor{aa}    11   & \cellcolor{aa} 13 \\
\rccc{aa}{aa}
\cellcolor{aa}
                                        & \cellcolor{aa}C. Out.: No      & \cellcolor{aa}       2 &  \cellcolor{aa} 3 &   \cellcolor{aa}1   &   \cellcolor{aa}        4 &  \cellcolor{aa}  6   &    \cellcolor{aa}     5   & \cellcolor{aa}       5 & \cellcolor{aa}        7   & \cellcolor{aa} 6   & \cellcolor{aa}    6   & \cellcolor{aa}     7   & \cellcolor{aa}  8   & \cellcolor{aa}       8   & \cellcolor{aa}10   &   \cellcolor{aa}      9   &  \cellcolor{aa}7   &   \cellcolor{aa}         10   & \cellcolor{aa} 10   & \cellcolor{aa}  11   &   \cellcolor{aa}   7   &  \cellcolor{aa}12   &    \cellcolor{aa}  13   &  \cellcolor{aa}13 \\
\rccc{aa}{aa}
\cellcolor{aa}
                                        & \cellcolor{aa}L.+C. Out.: Yes    &    \cellcolor{aa}     2 &   \cellcolor{aa}3 &   \cellcolor{aa}2   &     \cellcolor{aa}      5 &  \cellcolor{aa}  6   &   \cellcolor{aa}      6   & \cellcolor{aa}       6 & \cellcolor{aa}        11   & \cellcolor{aa} 6   & \cellcolor{aa}     5   & \cellcolor{aa}     9   & \cellcolor{aa}   6   & \cellcolor{aa}       7.5 &  \cellcolor{aa}9   & \cellcolor{aa}       10   & \cellcolor{aa}5   &    \cellcolor{aa}         8.5 &   \cellcolor{aa} 9   &    \cellcolor{aa}10   &   \cellcolor{aa}   4   & \cellcolor{aa}13   &    \cellcolor{aa}  10   & \cellcolor{aa}11 \\
  \rccc{aa}{aa}
\cellcolor{aa}
   \multirow{-6}*{Outage Experience}  & \cellcolor{aa}L.+C. Out.: No    &    \cellcolor{aa}     2 &   \cellcolor{aa}3 & \cellcolor{aa} 1   &      \cellcolor{aa}     4 &    \cellcolor{aa}6   &   \cellcolor{aa}      5   & \cellcolor{aa}       5 & \cellcolor{aa}         7   & \cellcolor{aa}6   & \cellcolor{aa}    6   & \cellcolor{aa}     7   & \cellcolor{aa}  8   &   \cellcolor{aa}      8   & \cellcolor{aa}10   &   \cellcolor{aa}      9   & \cellcolor{aa}6   &    \cellcolor{aa}         9.5 & \cellcolor{aa}  10   &  \cellcolor{aa}  12   &   \cellcolor{aa}   7   &  \cellcolor{aa}12   &  \cellcolor{aa}    14   & \cellcolor{aa}14 \\
   
  \hline 
  
  \rccc{ba}{ba}  
  \cellcolor{ba}
                                   &  \cellcolor{ba}Yes  &       \cellcolor{ba}2 &    \cellcolor{ba}3 &     \cellcolor{ba}3   &     \cellcolor{ba}    4 &   \cellcolor{ba}6   &    \cellcolor{ba}   5   &   \cellcolor{ba}    5 &   \cellcolor{ba}    11   &    \cellcolor{ba}6   &       \cellcolor{ba}5.5 &       \cellcolor{ba}13   &      \cellcolor{ba}7   &     \cellcolor{ba}  8   &   \cellcolor{ba}19.5 &       \cellcolor{ba}11   &    \cellcolor{ba}7.5 &       \cellcolor{ba}    3   &   \cellcolor{ba}15   &   \cellcolor{ba}12   &   \cellcolor{ba} 5.5 &    \cellcolor{ba}15.5 &       \cellcolor{ba}16.5 &    \cellcolor{ba}18 \\
   \rccc{ba}{ba}
     \cellcolor{ba}
     \multirow{-2}*{First Responder} &   \cellcolor{ba}No &          \cellcolor{ba}2 &    \cellcolor{ba}3 &     \cellcolor{ba}1   &       \cellcolor{ba}  4 &     \cellcolor{ba}6   &       \cellcolor{ba}5   &      \cellcolor{ba} 5 &         \cellcolor{ba}7   &    \cellcolor{ba}6   &     \cellcolor{ba}6   &   \cellcolor{ba} 7   &    \cellcolor{ba}8   &         \cellcolor{ba}8   &  \cellcolor{ba}10   &        \cellcolor{ba}10   &  \cellcolor{ba}6   &       \cellcolor{ba}   10   &    \cellcolor{ba}10   &    \cellcolor{ba}11   &      \cellcolor{ba}7   &  \cellcolor{ba}12   &     \cellcolor{ba}13   &  \cellcolor{ba}13 \\
  
  \hline 
  
\rccc{aa}{aa}
  \cellcolor{aa}
                                               & \cellcolor{aa}Low    &     \cellcolor{aa}   1 &   \cellcolor{aa}3 &  \cellcolor{aa}2   &        \cellcolor{aa}  4 &    \cellcolor{aa}5   &         \cellcolor{aa}5   &         \cellcolor{aa}4 &          \cellcolor{aa}7   &  \cellcolor{aa}5   &    \cellcolor{aa} 5   &     \cellcolor{aa}8   &     \cellcolor{aa}6   &       \cellcolor{aa} 7.5 &  \cellcolor{aa}10   &     \cellcolor{aa}  10   &  \cellcolor{aa}5   &      \cellcolor{aa}      8   &   \cellcolor{aa}13   &  \cellcolor{aa}  8   &      \cellcolor{aa}4   & \cellcolor{aa}14   &     \cellcolor{aa}12   &  \cellcolor{aa}13 \\
\rccc{aa}{aa}
 \cellcolor{aa}                                             & \cellcolor{aa}Medium       &      \cellcolor{aa}  2 &    \cellcolor{aa}3 &  \cellcolor{aa}2   &     \cellcolor{aa}     4 &   \cellcolor{aa}6   &         \cellcolor{aa}5   &       \cellcolor{aa} 5 &          \cellcolor{aa}7   &   \cellcolor{aa}6   &    \cellcolor{aa} 6   &  \cellcolor{aa}   7   &  \cellcolor{aa} 8   &     \cellcolor{aa}   8   &  \cellcolor{aa}10   &     \cellcolor{aa}  10   & \cellcolor{aa}6   &      \cellcolor{aa}      9   &    \cellcolor{aa}10   &   \cellcolor{aa}11   &   \cellcolor{aa}  6   &  \cellcolor{aa}12   &   \cellcolor{aa}  12.5 & \cellcolor{aa}13 \\
\rccc{aa}{aa} 
 \cellcolor{aa}
    \multirow{-3}*{Technology Affinity} &  \cellcolor{aa}High      &  \cellcolor{aa}      2 &  \cellcolor{aa}3 &   \cellcolor{aa}1   &    \cellcolor{aa}      5 &  \cellcolor{aa} 6.5 &    \cellcolor{aa}    6   &  \cellcolor{aa}      5 &   \cellcolor{aa}      8   &   \cellcolor{aa}7   &   \cellcolor{aa}  6   &  \cellcolor{aa}   6.5 &   \cellcolor{aa}7.5 &     \cellcolor{aa}   9   &  \cellcolor{aa}10   &    \cellcolor{aa}    9   &  \cellcolor{aa}9   &       \cellcolor{aa}    10   &     \cellcolor{aa}8   &     \cellcolor{aa}12   &      \cellcolor{aa}8   &   \cellcolor{aa}12   &    \cellcolor{aa} 14.5 &   \cellcolor{aa}15 \\
  
  \hline 
  
\end{tabular}
\end{table*}

\begin{table*}
  \caption{
      \captionheadline{Study Sample Description}
      This table describes our study sample in terms of the variables derived from the 20 questions of our questionnaire.
      We depict demographic variables from Q1~-~Q4 and control variables from the remaining questionnaire.
      Q6 and Q9 were simple confirmations, dependent variables Q11~-~Q13 answered our research questions (Section~\ref{sec: results}), and we used the open-ended questions Q18~-~Q20 to discuss data quality.
      We introduce categories for each variable to simplify discussions about individual groups.
      For numerical variables, we report the mean $\mu$, median $M$, and the \acf{IQR}.
  }
  \Description{
      Six columns: Variable, Category, Value, Goal, Sample, and Description.
      There are rows for the following variables: Age, Gender, Education, Income, Federal State, City Size, Residence Time, Screen Time, Crisis Experience, Outage Experience, First Responder, and Technology Affinity.
      There rows in the Description column align with the Variables.
      The remaining columns however are divided into categories for each variable, e.g., distinguishing younger, middle-aged, and older participants for the age variable.
  }
\label{tab:study-sample}
\small
\begin{tabular}{llrrrL{220pt}}
  \hline 
  
  \textbf{Variable}  & \textbf{Category} & \textbf{Value}    & \textbf{Goal} & \textbf{Sample}   & \textbf{Description}   \\
                              &                   &                   &               & ($N = \nfinal$)   &                       \\
  
  \hline 
  
\cc{aa}&        \cc{aa}& \cc{aa}\SIrange{18}{29}{\year}   & \cc{aa}\SI{21}{\percent} &\cc{aa}\SI{12}{\percent} (103)   & \cc{aa}\\
 \cc{aa}&\cc{aa}\multirow{-2}*{Younger}       &\cc{aa}\SIrange{30}{39}{\year}   & \cc{aa}\SI{19}{\percent} & \cc{aa}\SI{20}{\percent} (174)   & \cc{aa}\\
\cc{aa}&            \cc{aa}&\cc{aa}\SIrange{40}{49}{\year}   &\cc{aa}\SI{18}{\percent} &\cc{aa}\SI{21}{\percent} (180)   & \cc{aa}\\
   \cc{aa}& \cc{aa}\multirow{-2}*{Middle-Aged}   & \cc{aa}\SIrange{50}{59}{\year}   & \cc{aa}\SI{24}{\percent} &\cc{aa}\SI{26}{\percent} (225)   &\cc{aa}\cc{aa}\\
 \cc{aa}&               \cc{aa}& \cc{aa}\SIrange{60}{69}{\year}   &\cc{aa}\SI{19}{\percent} & \cc{aa}\SI{20}{\percent} (175)   & \cc{aa}\\
\cc{aa}\multirow{-6}*{Age (Q1)}  & \cc{aa}\multirow{-2}*{Older}         &\cc{aa}$\geq$\SI{70}{\year}      &\cc{aa}& \cc{aa}\SI{0}{\percent}    (0)   & \cc{aa}\\
  
  \hline 
  
\cc{ba}          & \cc{ba}Men          &\cc{ba}Male          &\cc{ba}\SI{50}{\percent} & \cc{ba}\SI{49}{\percent} (416)   & \cc{ba}\\
 \cc{ba}           &\cc{ba}Women        &\cc{ba}Female        &\cc{ba}\SI{50}{\percent} &\cc{ba}\SI{51}{\percent} (437)   & \cc{ba}\\
\cc{ba}\multirow{-3}*{Gender (Q2)}   &\cc{ba}Non-Binary    &\cc{ba}Non-Binary    &  \cc{ba}& \cc{ba}\SI{0}{\percent}    (4)   & \cc{ba}\\
  
  \hline 
  
\cc{aa}      &\cc{aa}Low       & \cc{aa}Low       &\cc{aa}\SI{26}{\percent} &\cc{aa}\SI{13}{\percent} (109)   & \cc{aa}\\
 \cc{aa}     &\cc{aa}Medium    & \cc{aa}Medium    &\cc{aa}\SI{33}{\percent} &\cc{aa}\SI{59}{\percent} (502)   & \cc{aa}\\
 \cc{aa}\multirow{-3}*{Education\mytablemark{1} (Q3)}   & \cc{aa}High      & \cc{aa}High      & \cc{aa}\SI{40}{\percent} & \cc{aa}\SI{29}{\percent} (246)   & \cc{aa}\\
  
  \hline 
  
\cc{ba}           &  \cc{ba}      & \cc{ba}<\SI{1000}{\EUR}              &\cc{ba}\SI{6}{\percent}  &\cc{ba}\SI{6}{\percent}   (48)   & \cc{ba}\\
 \cc{ba}& \cc{ba}\multirow{-2}*{Low}       &\cc{ba}\SIrange{1000}{1999}{\EUR}    & \cc{ba}\SI{16}{\percent} &\cc{ba}\SI{18}{\percent} (152)   & \cc{ba}\\
\cc{ba}            & \cc{ba}       &\cc{ba}\SIrange{2000}{2999}{\EUR}    & \cc{ba}\SI{22}{\percent} & \cc{ba}\SI{24}{\percent} (202)   & \cc{ba}\\
\cc{ba}&\cc{ba}\multirow{-2}*{Medium}    &\cc{ba}\SIrange{3000}{3999}{\EUR}    &\cc{ba}\SI{21}{\percent} &\cc{ba}\SI{22}{\percent} (186)   & \cc{ba}\\
\cc{ba}            & \cc{ba}       & \cc{ba}\SIrange{4000}{4999}{\EUR}    & \cc{ba}\SI{22}{\percent} & \cc{ba}\SI{16}{\percent} (135)   & \cc{ba}\\
\cc{ba}\multirow{-6}*{Income (Q4)}   &\cc{ba}\multirow{-2}*{High}      & \cc{ba}$\geq$\SI{5000}{\EUR}         &\cc{ba}\SI{14}{\percent} &\cc{ba}\SI{14}{\percent} (122)   & \cc{ba}\\
  
  \hline 
  
 \cc{aa}Federal State (Q5)    & \cc{aa}&   \cc{aa}&  \cc{aa}& \cc{aa}&  \cc{aa}Our sample covers all 16 federal states\mytablemark{2}. \\
  
  \hline 
  
 \cc{ba}              &  \cc{ba}Smaller      & \cc{ba}\numrange{100000}{199999} &   \cc{ba}&  \cc{ba}\SI{13}{\percent} (138)   & \cc{ba}\\
  \cc{ba}               & \cc{ba}Medium       & \cc{ba}\numrange{200000}{499999} &  \cc{ba}&  \cc{ba}\SI{22}{\percent} (201)   & \cc{ba}Our sample covers 79 of the 82 German \\
  \cc{ba}               &  \cc{ba}Larger       &  \cc{ba}\numrange{500000}{999999} &   \cc{ba}&  \cc{ba}\SI{29}{\percent} (244)   & \cc{ba}major cities \\
    \cc{ba}\multirow{-4}*{City Size (Q7)}    &  \cc{ba}Metropolis   & \cc{ba}$\geq$\num{1000000}       &   \cc{ba}& \cc{ba}\SI{32}{\percent} (274)   & \cc{ba}\\
  
  \hline 
  
 \cc{aa}  & \cc{aa}Short     &  \cc{aa}<4 years          &  \cc{aa}& \cc{aa}\SI{13}{\percent} (109)   & \cc{aa}\\
    \cc{aa} & \cc{aa}Medium    & \cc{aa}4-9 years         &    \cc{aa}&  \cc{aa}\SI{59}{\percent} (502)   & \cc{aa}\\
 \cc{aa}\multirow{-3}*{Residence Time (Q8)}   & \cc{aa}Long      & \cc{aa}$\geq$10 years    &  \cc{aa}&  \cc{aa}\SI{29}{\percent} (246)   & \cc{aa}
  \multirow{-3}*{$\mu=13.48$; $M=16$; $\text{IQR}=3$} \\
  
  \hline 
  
 \cc{ba}              & \cc{ba}Low      & \cc{ba}<\SI{3}{\hour}        &\cc{ba}& \cc{ba}\SI{41}{\percent} (348)   & \cc{ba}\\
   \cc{ba}               & \cc{ba}Medium   & \cc{ba}\SIrange{3}{5}{\hour} &\cc{ba}& \cc{ba}\SI{31}{\percent} (266)   & \cc{ba}\\
   \cc{ba}\multirow{-3}*{Screen Time (Q10)} &  \cc{ba}High     & \cc{ba}$\geq$\SI{5}{\hour}   & \cc{ba}&\cc{ba}\SI{28}{\percent} (243)   & \cc{ba}
  \multirow{-3}*{$\mu=\text{\hourmin{3}{44}}$; $M=\SI{3}{\hour}$; $\text{IQR}=\SI{3}{\hour}$} \\
  
  \hline 
  
\cc{aa}     &\cc{aa}Earthquake    &  \cc{aa}& \cc{aa}& \cc{aa}\SI{14}{\percent} (122)   & \cc{aa}\\
 \cc{aa}& \cc{aa}Flooding      &\cc{aa}&\cc{aa}&\cc{aa}\SI{19}{\percent} (163)   & \cc{aa}\\
\cc{aa}      & \cc{aa}Storm         &  \cc{aa}&  \cc{aa}& \cc{aa}\SI{49}{\percent} (424)   & \cc{aa}\\
  \cc{aa}      &\cc{aa}Wildfire      &  \cc{aa}& \cc{aa}&\cc{aa}\SI{5}{\percent}   (46)   & \cc{aa}\\
  \cc{aa}\multirow{-5}*{Crisis Experience (Q14)}   &\cc{aa}War           & \cc{aa}&  \cc{aa}&\cc{aa}\SI{3}{\percent}   (24)   & \cc{aa}\\
  
  \hline 
  
\cc{ba}                               &   \cc{ba}L Outage       &\cc{ba}$\geq$\SI{6}{\hour}   &\cc{ba}& \cc{ba}\SI{25}{\percent} (213) & \cc{ba}\\
 \cc{ba}                              & \cc{ba}C. Outage       &\cc{ba}$\geq$\SI{6}{\hour}   &\cc{ba}& \cc{ba}\SI{14}{\percent} (116) & \cc{ba}\\
\cc{ba}\multirow{-3}*{Outage Experience\mytablemark{3} (Q15)} &\cc{ba}L.+C. Outage    & \cc{ba}$\geq$\SI{6}{\hour}   &  \cc{ba}& \cc{ba}\SI{13}{\percent} (113) & \cc{ba}\\
  
  \hline 
  
  \cc{aa}First Responder (Q16) &   \cc{aa}&  \cc{aa}&   \cc{aa}&  \cc{aa}\SI{3}{\percent} (28) & \cc{aa}\\
  
  \hline 
  
\cc{ba}                     & \cc{ba}Low       & \cc{ba}\numrange{1}{2.5}     & \cc{ba}& \cc{ba}\SI{12}{\percent} (100)   & \cc{ba}\\
  \cc{ba}                      &\cc{ba}Medium    &\cc{ba}\numrange{2.5}{4.5}   &\cc{ba}&\cc{ba}\SI{67}{\percent} (578)   & \cc{ba}\\
  \cc{ba}\multirow{-3}*{Technology Affinity (Q17)} & \cc{ba}High      & \cc{ba}\numrange{4.5}{6}     & \cc{ba}& \cc{ba}\SI{21}{\percent} (179)   & \cc{ba}
  \multirow{-3}*{$\mu=3.68$; $M=3.67$; $\text{IQR}=1.33$; $\alpha=.9$} \\
  
  \hline 
  \mytablenote{1}{6}{\textwidth}{"Low" is \surveyansweroptions{"None (yet)," "Hauptschulabschluss"}, "Medium" is \surveyansweroptions{"Polytechnische Oberschule," "Mittlere Reife, Realschulabschluss," "(Fach-)Hochschulreife"}, "High" is \surveyansweroptions{"Bachelor," "Master," "Diplom,"  "Promotion"}.}
  \mytablenote{2}{6}{\textwidth}{\SI{6}{\percent} \ac{BW}, \SI{11}{\percent} \ac{BY}, \SI{16}{\percent} \ac{BE}, \SI{1}{\percent} \ac{BB}, \SI{2}{\percent} \ac{HB}, \SI{8}{\percent} \ac{HH}, \SI{6}{\percent} \ac{HE}, \SI{1}{\percent} \ac{MV}, \SI{5}{\percent} \ac{NI}, \SI{28}{\percent} \ac{NW}, \SI{3}{\percent} \ac{RP}, \SI{0}{\percent} \ac{SL}, \SI{7}{\percent} \ac{SN}, \SI{3}{\percent} \ac{ST}, \SI{3}{\percent} \ac{SH}, and \SI{2}{\percent} \ac{TH}.}
    \mytablenote{3}{6}{\textwidth}{"L." and "C." are abbreviations for "Landline" and "Cellular," respectively.}
\end{tabular}
\end{table*}

\end{document}